\documentclass[%
preprint,
nofootinbib,
 amsmath,amssymb,
 aps,
]{revtex4-2}

\usepackage{graphicx}
\usepackage{dcolumn}
\usepackage{bm}

\newcommand{\apss}{Ap\&SS}
\newcommand{\aap}{A\&A}
\newcommand{\apjl}{ApJ Lett.}
\newcommand{\mnras}{MNRAS}

\usepackage{footmisc}

\pdfoutput=1
\begin{document}

\preprint{APS/123-QED}

\title{Stellar structure, magnetism and the variational principle}
\author{A. \v Cade\v z}
 \email{andrej.cadez@fmf.uni-lj.si}
\author{A. Mohori\v c}%
\affiliation{ 
Department of Physics, University of Ljubljana, Ljubljana, Slovenia
}%

\author{ M. Calvani}
\affiliation{%
Retired from INAF,
Astronomical Observatory of Padova,
Padova, Italy
}


\date{\today}

\begin{abstract}
  Matter interacts with two fundamental long range forces: gravity and electromagnetism. While elementary building blocks of matter always contribute to the gravitational potential, the electromagnetic vector potential was expected to cancel out in large systems because of the symmetry between positive and negative charges. Yet, long range electromagnetic phenomena are present in astrophysical systems, so that the cancellation can not be considered as perfect.  The aim of this paper is to find a possible model for stationary aggregation of matter to make a ``star" which consistently includes angular momentum and electromagnetic phenomena. We recast the standard polytropic stellar model as a variational problem and extend it to include the kinetic energy of (rigid) rotation and the electromagnetic energy of interaction between positively and negatively charged baryonic matter. We argue that the electromagnetic contribution to the action should be represented by the minimal energy needed to induce the stellar magnetic dipole moment. This energy consists of two parts: the pure electromagnetic energy, which can be expressed as a surface integral, and the free energy difference between the magnetized and unmagnetized state of matter. This second part of the action is formulated by considering the magnetized quantum state of a Fermi gas of electrons in a sea of cold ions. Differential forms calculus is a useful tool to perform the mathematical analysis (\cite{Cecile1996}). Our mathematical model includes electromagnetism to stellar modeling in a way that is consistent with the linearized version of general relativity. The complete model  presents a complex open boundary problem. The problem can be solved exactly under simplifying circumstances. The distribution of stellar objects in the phase diagram, based on the simplified model, exhibits a pattern which may prompt research to provide a deeper understanding of the natural balance between matter, gravitation, and electromagnetism.

\end{abstract}

\maketitle
\section{Introduction}

The equations of stellar structure, which include the interaction of matter with the gravitational field, the production of energy by nuclear reactions, and the transport of energy via convection and radiation, have been amazingly successful in explaining the observed properties of stars and describing their history from birth to death. The long-range effects of the electromagnetic force are traditionally neglected, since deviations from spherical symmetry caused by stellar rotation and magnetic fields appear to have minimal effects on stellar evolution. Yet, stars and other celestial objects do spin and exhibit magnetic properties. The magnetism of pulsars seems to be the most interesting property of neutron stars, while the magnetism of the Sun, Earth, and planets presents us with a number of unanswered questions.
This paper presents a Hamiltonian action that leads to the basic equations of stellar structure through a variational principle. The action includes stellar rotation and magnetism in a straightforward way and leads to a phase diagram in which celestial objects are sorted according to the relative strength of their angular momentum and magnetic field.
The plan of the paper is as follows: Section \ref{LaneEmden1} recapitulates the derivation of the Lane-Emden equation for a rotating polytrope and expresses the energy released during the process of forming a star from infinitely rarefied gas in its ground state. This energy forms an action that, together with suitable constraints, leads back to the Lane-Emden equation via Lagrange equations. Section \ref{SSec3_1} introduces a brief formulation of Maxwell’s equations in a form convenient for the analysis of different configurations of currents driving the magnetic field.  The current distribution producing a magnetic moment with the minimum amount of energy is introduced. Section \ref{SSec3_2} treats the interaction of a degenerate gas of electrons with the electromagnetic field. In section \ref{MagnLaEm} The electromagnetic energy is added to the action, which leads to generalized Lane-Emden equations that include the contribution of a stellar magnetic dipole moment. A subsection \ref{MGSI} is devoted to the fact that magnetic action is expressed by a surface integral and to the instability of this surface in the case of a strong magnetic field. In section \ref{RotMagSt}, we introduce rotation and develop a model of a rotating magnetic star and classify a number of stars according to this model.  

\section{Rotating stars and Lane-Emden equation through variational principle}
\label{LaneEmden1}
Polytropic models, described by the Lane-Emden equation, are very simple stellar models. They are particularly useful in describing the mass-radius relation of compact stars thus providing a tool to understand the state of degenerate matter in observed stellar objects. 

Lane-Emden equation follows from: local static equilibrium, the law of gravitation and  the polytropic equation of state. We note  that this equation can also be derived from a variational principle minimizing the total energy of the star which includes the enthalpy of stellar matter (the work done by gravitation to compress matter below the surface of the star) and the gravitational energy. Since stars exhibit both rotation and magnetism, it seems natural that global effects of angular momentum and magnetism could also be described by a generalized Lane-Emden equation that would follow from a variational principle minimizing the energy which includes also rotational and magnetic energy. 
The equations leading to  the polytropic equation for a rotating star are\footnote{$\nabla$ stands for gradient and $\Delta$ for Laplacian operator.}:
\begin{align}
&{\rm polytropic\quad equation\quad of\quad state}  &p\quad &=\quad K\,\rho^{1+1/n}\, ,\nonumber \\
&{\rm local\quad equilibrium}\qquad &\rho\frac{d\vec v}{dt}&=-\rho \nabla \Phi_g - \nabla p \, ,\label{LocEqil}\\
&{\rm gravitational\quad field}\qquad &\Delta\Phi_g&=\quad 4\pi G \rho \, ,\nonumber
\end{align}
together with the constraint that the velocity field is that of  rigid rotation: $\vec v=\vec \omega\times \vec r$ and $\frac{d\vec v}{dt}=\vec \omega \times \left (\vec \omega \times \vec r\right)=\nabla \left(-\frac{1}{2}\omega^2R^2\right)$. Here $R$ is the distance from the axis of rotation. Taking into account also the equation of state and dividing  the local equilibrium equation by $\rho$, one obtains $
\nabla \left(\Phi_g-\frac{1}{2}\omega^2R^2+(n+1)K\rho^{\frac{1}{n}}\right)=0
$ which is a constraining equation resulting from the requirement of rigid rotation. Since $p$ and $\rho$ are directly related, it is customary to introduce a new field $\Theta$ such that $\rho=\rho_0 \Theta^n$ and $p=p_0 \Theta^{n+1}$, which brings the constraining equation in the simple form:
\begin{equation}
\Phi_g=-(n+1)\frac{p_0}{\rho_0}\Theta+\frac{1}{2}\omega^2R^2+\beta\, ,
\label{RotConstr}
\end{equation} 
where $\beta$ is an integration constant. 
The Lane-Emden equation follows by applying the gravitational field equation:
\begin{equation}
\frac{n+1}{4\pi G \rho_0}\frac{p_0}{\rho_0}\Delta \Theta+\Theta^n-\frac{\omega^2}{2\pi G \rho_0}=0\, .
\label{LErot}
\end{equation}

The boundary conditions on $\Theta$ follow from the general requirement that the divergence of  stress energy tensor belonging to all interacting fields and matter vanishes everywhere in space. The Lane-Emden problem for rotating polytropes is thus an open boundary problem. 
At the surface of the star  matter density and pressure vanish, but gravitation extends beyond the surface, therefore the vanishing of stress-energy divergence at the surface requires the gravitational field and its derivative to be continuous at the boundary - the surface of the star. 

In order to formulate the problem on the basis of  variational principle, we write down the total energy of the star, i.e. all energies that have been released or have done work during the process of condensation of stellar matter from the infinitely rarified state to the current state of stable stellar equilibrium. Gravitational energy can be expressed in the following forms:
\begin{align}
W_g&=\frac{1}{2}\displaystyle \int_V \rho\, \Phi_g dV=\frac{1}{8\pi G}\displaystyle \int_V \Phi_g\,\Delta \Phi_g\, dV\nonumber\\&=-\frac{1}{8\pi G}\displaystyle \int_V \nabla \Phi_g \cdot \nabla \Phi_g \, dV+\frac{1}{8\pi G}\displaystyle \int_{\partial V} \Phi_g \nabla \Phi_g\cdot d\vec S\,  ,
\label{GravEn}
\end{align} 
kinetic energy of rotation as:
\begin{align}
W_{rot}&=\displaystyle \int_V \frac{1}{2}\rho \,\omega^2 R^2 \,dV=\frac{\omega^2}{8\pi G}\displaystyle \int_V R^2\Delta \Phi_g\, dV\nonumber\\
&= -\frac{\omega^2}{4\pi G}\displaystyle \int_V \vec R\cdot \nabla \Phi_g\, dV+\frac{\omega^2}{8\pi G}\displaystyle \int_{\partial V} R^2\nabla \Phi_g\cdot d\vec S\, ,
\label{RotEn}
\end{align}
and internal energy of stellar matter\footnote{A polytropic model assumes the entropy per particle to be independent of position, so that  $\int_V T\, dS=W_S$ is a given constant for a given model. This condition is a constraint of polytropic model.}:
\begin{equation}
W_{H}=\displaystyle \int_V p\, dV-\displaystyle \int_V T\, dS\quad\quad \longrightarrow\quad\quad\displaystyle \int_V p_0\, \Theta^{n+1}\,dV\, .
\label{FreeEn}
\end{equation}
Here $\int_V$ represents integral over the volume and $\int_{\partial V}$  integral over the outer surface of the star. Note that surface integrals result from volume integrals over exterior region, which  contain integrands in the form of divergences  going to zero at spatial infinity.    
 
The action for the corresponding variational principle is the sum of all three energies taking into account the rigid rotation constraint (\ref{RotConstr}) plus the mass ($M=\frac{1}{4\pi G} \int_{\partial V}\nabla \Phi_g\cdot d\vec S$) and angular momentum ($L=\frac{2}{\omega}W_{rot}$) constraints with Lagrange multipliers $\lambda_M$ and $\lambda_L$: 

\footnotesize{
\begin{align}
A=&-\displaystyle \int_V p_0\left( \frac{(n+1)^2p_0}{8\pi G \rho_0^2}\nabla \Theta \cdot \nabla \Theta-\frac{(n+1)( \lambda_L+\omega)\omega}{2\pi G \rho_0}\vec R\cdot \nabla \Theta\right .\nonumber\\&\left .\qquad\qquad\qquad\qquad -\Theta^{n+1}+(\lambda_L+\frac{3\omega}{4})\frac{\omega^3}{2\pi G p_0}\vec R\cdot \vec R \right)\, dV\nonumber \\
&- \frac{1}{\omega}\displaystyle \int_{\partial V}\frac{(n+1)p_0}{8\pi G } \left[\frac{p_0}{\rho_0}\left( 4(n+1)\lambda_L + (4+3n) \omega \right)\Theta \nabla \Phi_g\right .\nonumber\\   &\qquad\left .-2 \left( 2(\lambda_L+\frac{3}{4}\omega)\Phi_g - (2\lambda_L+\omega)\beta+\omega \lambda_M\right)\nabla \Phi_g \right]\cdot d\vec S\, .
\end{align}}
The field equation (\ref{LErot}) follows from the volume part of the above action for $\lambda_L=-\omega /2$. The surface part of action simplifies, because $\Theta_{\partial V}=0$, so  the first term  in square brackets vanishes and the surface action becomes $ \displaystyle \int_{\partial V}\frac{(n+1)p_0}{8\pi G } \left[(\Phi_g+2 \lambda_M)\nabla \Phi_g \right ]\cdot d\vec S $.  The constraints are satisfied if $-2\lambda_M$ is  the  average gravitational potential on the surface.  

\subsection{On solution of Lane-Emde equation }
\label{SSec2_1}
The first effect of rotation of a star on its shape is flattening - equatorial radius becomes larger than polar radius.  The condition (\ref{RotConstr}) requires the gravitational potential on the surface of the star to be $\left(\frac{1}{2}\omega^2\, R^2+\beta\right)$, so that for slow rotation the dominant term breaking the spherical symmetry is the quadrupole. This suggests that  elliptical shape is a good approximation for slowly rotating stars. 

To illustrate the mechanism, we recapitulate the derivation of  the classical Maclaurin model of gravitating incompressible fluid, which is the polytropic model with $n=0$. 
Equation \ref{LErot} takes the very simple form $\frac{1}{4\pi G \rho}\frac{p_0}{\rho_0}\Delta \Theta+1-\frac{\omega^2}{2\pi G \rho_0}=0$, which upon the introduction of constants $\Omega^2=2\pi G \rho_0$, $a^2=\frac{p_0}{2\rho_0\, \Omega^2}$, $\tilde \omega=\omega/\Omega$ and $\tilde \Delta=a^2\Delta$ becomes
\begin{equation}
1-{\tilde \omega}^2+\tilde \Delta \Theta=0 \, .
\label{LEn0}
\end{equation} 
Since the surface of the star, where $\Theta=0$, is expected to be elliptical,  we seek the solution in elliptic coordinates $\lbrace \xi,\, \theta,\, \phi\rbrace$ defined by:  
\begin{equation}
\begin{pmatrix}x\\y\\z\end{pmatrix}=\begin{pmatrix}a\sqrt{\xi^2+e^2}\sin \theta\cos \phi\\a\sqrt{\xi^2+e^2}\sin \theta\sin \phi\\a\,\xi \cos \theta\end{pmatrix}\, .
\label{EliptCOO}
\end{equation}
The Laplacian in these coordinates is:
\begin{align}
\tilde \Delta&=\frac{1}{\xi^2+e^2\cos^2\theta}\left[\frac{\partial}{\partial \xi}\left((\xi^2+e^2)\frac{\partial}{\partial \xi}\right)\frac{1}{\sin\theta}\left(\sin \theta \frac{\partial }{\partial\theta}\right)\right] 
+\frac{1}{(\xi^2+e^2)\sin^2\theta}\frac{\partial^2 }{\partial \phi^2}\, .\nonumber
\end{align}
The complete set of cylindrically symmetric solutions of Laplace's equation is:
\begin{align}
\psi_l^{(int)}&=P_{l}(i\frac{\xi}{e})\, P_l(\cos \theta) &0<\xi<1\, ,\nonumber \\
\psi_l^{(ext)}&=\left(i Q_l(i\frac{x}{e})+\frac{\pi}{2}P_{l}(i\frac{\xi}{e})\right)\, P_l(\cos \theta) &1<\xi<\infty\, ,
\label{ExtLapl}
\end{align}
where $P_l(x)$ and $Q_l(x)$ are Legendre polynomials and Legendre functions of order $l$ for $l=0,\,1,\,2\dots \infty$.

The  solution of eq.(\ref{LEn0}) satisfying the boundary condition $\Theta=0$ at the surface $\xi=1$ is:
\begin{equation}
\Theta(\xi,\theta)=\frac{1-{\tilde \omega}^2}{2(3+e^2)}(1-\xi^2)(1+e^2\,\cos^2\theta)\, .
\end{equation}
The gravitational potential inside the star follows from eq.(\ref{RotConstr}) and becomes:
\begin{align}
&\Phi_g^{(int)}=\beta -\frac{p_0}{2\rho_0}\left [\frac{1-{\tilde \omega}^2}{(3+e^2)}(1-\xi^2)(1+e^2\,\cos^2\theta)-\frac{1}{2}{\tilde \omega}^2(\xi^2+e^2)\sin^2\theta\right]\, .
 \label{FiGNot}
\end{align}
The constant $\beta $ and the relation between $e $ and $\omega$ follow after the internal gravitational potential is smoothly joined to the external gravitational potential $\Phi_g^{(ext)}=c_0\,\psi_0^{(ext)}+c_2\,\psi_2^{(ext)}$at  $\xi=1$ (eq.\ref{ExtLapl}) written explicitly as:
\begin{equation}
\Phi_g^{(ext)}=c_0\, \,\arctan(\frac{e}{\xi})+c_2\,\left((3\xi^2+e^2)\,\arctan(\frac{e}{\xi})-3 e\,\xi
\right)P_2(\cos \theta)
\label{FiGZun}\, .
\end{equation}
The boundary conditions  $\Phi_g^{(ext)}=\Phi_g^{(int)}$ and  $\frac{d\Phi_g^{(ext)}}{d\xi}=\frac{d\Phi_g^{(int)}}{d\xi}$ at $\xi=1$ lead to:
\begin{align}
{\tilde \omega}^2&=\frac{1}{e^3}\left((3+e^2)\, \arctan(e)-3e\right)\nonumber\\ &\qquad  \rightarrow \frac{4}{15}e^2-\frac{8}{35}e^4+\frac{4}{21}e^6+\dots\, ,  \label{omegaepsilon} \\
\beta&=\frac{p_0}{2\rho_0 e^3}(1+e^2)\left(e-(1+e^2) \arctan(e)\right)\nonumber \\ & \qquad
\rightarrow\frac{p_0}{2\rho_0}\left(-\frac{2}{3}-\frac{8}{15}e^2+\frac{8}{105}e^4-\frac{8}{315}e^6+\dots\right)\, ,\nonumber \\
c_0&=-\frac{p_0}{\rho_0}\frac{1+e^2}{3 e}\, ,\nonumber \\
c_2&=-\frac{p_0}{\rho_0}\frac{1+e^2}{6 e^3}\, .\nonumber \\
\end{align}
This is equivalent to MacLaurin solution if $\varepsilon_{Mcl}^2=\frac{e^2}{1+e^2}$, (\cite{1965MNRAS.131...13M}) which is an exact solution (even if not dynamically stable for large $e$). 

This solution can be also obtained directly from variational principle for total energy, if one starts with the premise that a rotating self-gravitating incompressible fluid takes the shape of a rotational ellipsoid, and the only unknown is the relation between the angular momentum and ellipticity of the body. Gravitational energy and rotational energy can be readily computed, while the internal energy of the incompressible fluid is constant, independent of shape. So, we have:
\begin{align}
W_g\quad&=-\frac{3G\,M^2}{5R_{ef}}\frac{(1+e^2)^{1/3}}{e}\arctan e \nonumber \\ &{\underset{e\rightarrow 0}{\longrightarrow}}-\frac{3G\,M^2}{5R_{ef}}\left(1-\frac{1}{45}e^4+\frac{64}{2835}e^6+\dots\right)\, ,\nonumber \\
 W_{rot}&=\frac{1}{5}M\,R_{ef}^2\Omega^2{\tilde \omega}^2(1+e^2)^{1/3} \, ,
\label{RotHamPrinc}
\end{align}
where the effective radius is defined so that $\frac{4\pi}{3}R_{ef}^3$ is the volume of the fluid. The action $A(e_,{\tilde \omega})=W_g+(1+\lambda_L\frac{2}{\tilde\omega})W_{rot}$ has a minimum for a given angular momentum $\Gamma=\frac{2}{5}M\,R_{ef}^2\Omega \,{\tilde\omega\,(1+e^2)^{1/3}}$. The relation between $\tilde \omega$ and $e$ leading from this requirement is precisely (\ref{omegaepsilon}).
  
The incompressible model is simple and leads to an exact solution of the problem,  because the differential operator in eq.(\ref{LErot}) allows separation of variables  in such a way that angular functions  ($P_l(\cos \theta)$) are the same in the region inside the star and outside.   This makes it possible to exactly join the internal and external gravitational field on the surface for all values of the parameter $e$.  Other rotating polytropes do not share this property. In general, their solutions may only be calculated numerically and one must take into account the possibility that elliptic shape does not lead to the minimum energy solution. However, in the limit of slow rotation, polytropic models do converge to ellipsoidal shape with small eccentricity. The first coefficient in relation (\ref{omegaepsilon})  ($\frac{4}{15}\approx\left(\frac{\tilde \omega}{e}\right)^2$) is an increasing function of the polytropic index. In the limit $n=5$, when the mass is concentrated at the center of the star, it becomes $\frac{2}{3}$, while for the last marginally stable polytrope $n=3$ this coefficient is $0.65$. In other words, hard polytropes are more deformed with respect to $\tilde \omega$ than soft polytropes, but the factor $\left(\frac{\tilde \omega}{e}\right)^2$ between the hardest and softest is less than $3$, however one must remember that the radius of a soft polytropes with a given mass is so much larger, therefore, as a consequence of decreasing their $\Omega$, soft polytropes become unstable at smaller $\omega$.

\section{On electromagnetic action}
\label{SSec3}

Stellar magnetism may be a complicated phenomenon, especially if one would like to understand the variety of magnetic phenomena on our Sun. However, the common manifestation of stellar magnetic field is through the magnetic dipole, which is usually expressed as ``the magnetic field at  magnetic poles''. From this perspective the magnetic dipole field appears to be the principal component of stellar magnetism entering in the balance between gravitation, pressure and rotation.  

The extension of polytropic model to magnetism requires the addition of two energies to the variational action: a) magnetic dipole energy b) the energy of elementary magnetic dipoles within matter interacting with magnetic field. 
\subsection{Magnetic dipole energy}
\label{SSec3_1}
Magnetic dipole energy should,  according to  previous discussion, be the energy needed to establish the given (measured) magnetic dipole moment. We express this energy through the work done by currents inside the magnet volume ($V_m$) as:
$
W_{EM}=-\displaystyle \int_{-\infty}^{t}\int_{V_m}\vec j\cdot \vec E\, dV\, dt\, ,
$  
which can be transformed using Maxwell's equations\footnote{The following form is used:$\nabla\cdot\vec B=0$, $\nabla\cdot\vec D=\rho$, $\nabla\times \vec E=-\frac{\partial \vec B}{\partial t}$, $\nabla\times \vec H=\vec j+\frac{\partial \vec D}{\partial t}$, where $\vec B=\mu \mu_0 \vec H$ and $\vec D=\varepsilon \varepsilon_0 \vec E$. The fields $\vec B$ and $\vec E$ can also be replaced by vector ($\vec A$) and scalar ($\Phi$) potential: $\vec B=\nabla\times \vec A$, $\vec E=-\nabla \Phi-\frac{\partial \vec A}{\partial t}$} and vector identities as follows:
\footnotesize{
\begin{align}
W_{EM}=&\displaystyle \int_{-\infty}^{t}\int_{V_m}\left(-\nabla \times \vec H+\frac{\partial \vec D}{\partial t}\right)\cdot \vec E\, dV\, dt\nonumber\\
=&\displaystyle \int_{-\infty}^{t}\int_{V_m}\left[-\nabla\cdot\left(\vec H\times \vec E\right)-(\nabla \times \vec E)\cdot \vec H+\frac{\partial \vec D}{\partial t}\cdot \vec E\right]dV\, dt\nonumber\\
=&\displaystyle \int_{-\infty}^{t}\int_{\partial V_m}\left(\vec E\times \vec H\right)\cdot d\vec S\, dt +\displaystyle \int_{-\infty}^{t}\int_{V_m}\left(\frac{\partial \vec B}{\partial t}\cdot \vec H+\frac{\partial \vec D}{\partial t}\cdot \vec E\right)dV\, dt\, .
\label{EMWork}
\end{align}}
The surface integral in the above expression represents the Poynting flux propagating from the outer surface of the magnet and building the outer electromagnetic field by a wave spreading in outer space, as the interior field is slowly rising to its final value.  Evaluating this integral, it is convenient to express EM field with vector potential in radiation gauge, defined so that $\vec E=-\frac{\partial \vec A}{\partial t}$ and $\vec H=\frac{1}{\mu_0}\nabla \times \vec A$. The wave, by which magnetic field propagates in outer space, satisfies the wave equation and the boundary condition $\vec A(\vec r_s,t)=\vec A_S(\vec r_s)f(t)$ on the surface of the magnet. Here $f(t)$ is a slowly rising function from 0 at $t_0$ to 1 at $t_f$.  At final time $t_f$ the wave has spread to the distance $c(t_f-t_0)$ and is in any finite volume indistinguishable from the magnetic field of a static magnet if $t_0\rightarrow -\infty$. The surface integral evaluates to:
 \begin{align}W_S =&\displaystyle \int_{-\infty}^{t}\int_{\partial V_m}\left(\vec E\times \vec H\right)\cdot d\vec S\, dt=-\frac{1}{\mu_0}\displaystyle \int_{\partial V_m} \vec A_S\times(\nabla \times \vec A_S)\cdot d\vec S 
\displaystyle \int_{t_0}^{t_f} f'(t)f(t)\,dt\nonumber \\&=-\frac{1}{2\mu_0}\displaystyle \int_{\partial V_m} \vec A_S\times(\nabla \times \vec A_S)\cdot d\vec S\, .\label{MagSurfInt} \end{align}
The volume integral  in eq.{\ref{EMWork}} appears simple\footnote{Magnetic energy in exterior region is $W_e=\frac{1}{2\mu_0}\displaystyle \int_{V_e}\vec B\cdot\vec B\, dV=\frac{1}{2\mu_0}\displaystyle \int_{V_e}(\nabla \times \vec A)\cdot (\nabla \times \vec A) dV$. Use vector identity $(\nabla \times \vec A)\cdot (\nabla \times \vec A) =\nabla\cdot(\vec A\times(\nabla \times\vec A))-\vec A\cdot \Delta \vec A+\vec A\cdot \nabla(\nabla\cdot \vec A)$, note that  $\Delta\vec A=\frac{1}{c^2}\frac{\partial^2\vec A}{\partial t^2}$   in exterior region and $\nabla\cdot \vec A=0$ in radiation gauge.  $W_e$ consists of two terms, the first is a volume integral of a divergence, which can be transformed to the surface integral $W_S$ and the second term $-\frac{1}{c^2}\int_{V_e}\vec A\cdot \frac{\partial^2\vec A}{\partial t^2}\, dV$  can be made arbitrarily small in the limit $t_0\rightarrow \infty $.}, if the process of magnetic field generation can be considered as adiabatic and the  parameters  $\mu$ and $\varepsilon $ are constants. The result is simply: $\frac{1}{2\mu_0\mu}\int_V (B^2+\frac{\mu\,\varepsilon}{c^2}E^2)\, dV$. 
In this case magnetic energy can be written as:
\begin{equation}
W_{EM}=-\frac{1}{2\mu_0}\displaystyle \int_{\partial V} \vec A\times(\nabla \times \vec A)\cdot d\vec S+\frac{\mu_0}{2}\displaystyle \int_V \vec H\cdot (\vec H+\vec M)\, dV\, ,
\label{magnEn}
\end{equation}
where $\vec M$ is the contribution of internal polarization to magnetic field ($\vec B=\mu_0(\vec H+\vec M)$).

\newpage
At this point one should remember that the phenomenological parameter $\mu$ describes the amount of magnetization that matter contributes to magnetic
field, if matter is in a steady equilibrium with a given magnetic environment characterized by $\vec H$. However,  magnetization can not instantly follow  magnetic field, because  elementary charges, providing magnetizing currents,  are subject to their own  laws of  dynamics. Thus, during the process of alignment of magnetization with external magnetic field $\mu$ is not a constant and the interaction of the two fields may exchange energy with the system of elementary charges. The energy transferred to the system of charges is part of the volume integral in eq.(\ref{magnEn}) but can only be calculated if the polarizability including details of electromagnetic interaction between magnetic field and the system of charges is fully taken into account.  The assumption $\mu=\mathrm{const}$ neglects work needed to polarize matter. The result (\ref{magnEn}) thus  represents full magnetic energy, but the work needed to magnetize a magnet must include also  the difference between the free energy of polarized and unpolarized matter. The work needed to adiabatically polarize degenerate matter is calculated in the next section. 
\begin{figure}[h]
\includegraphics[width=9cm]{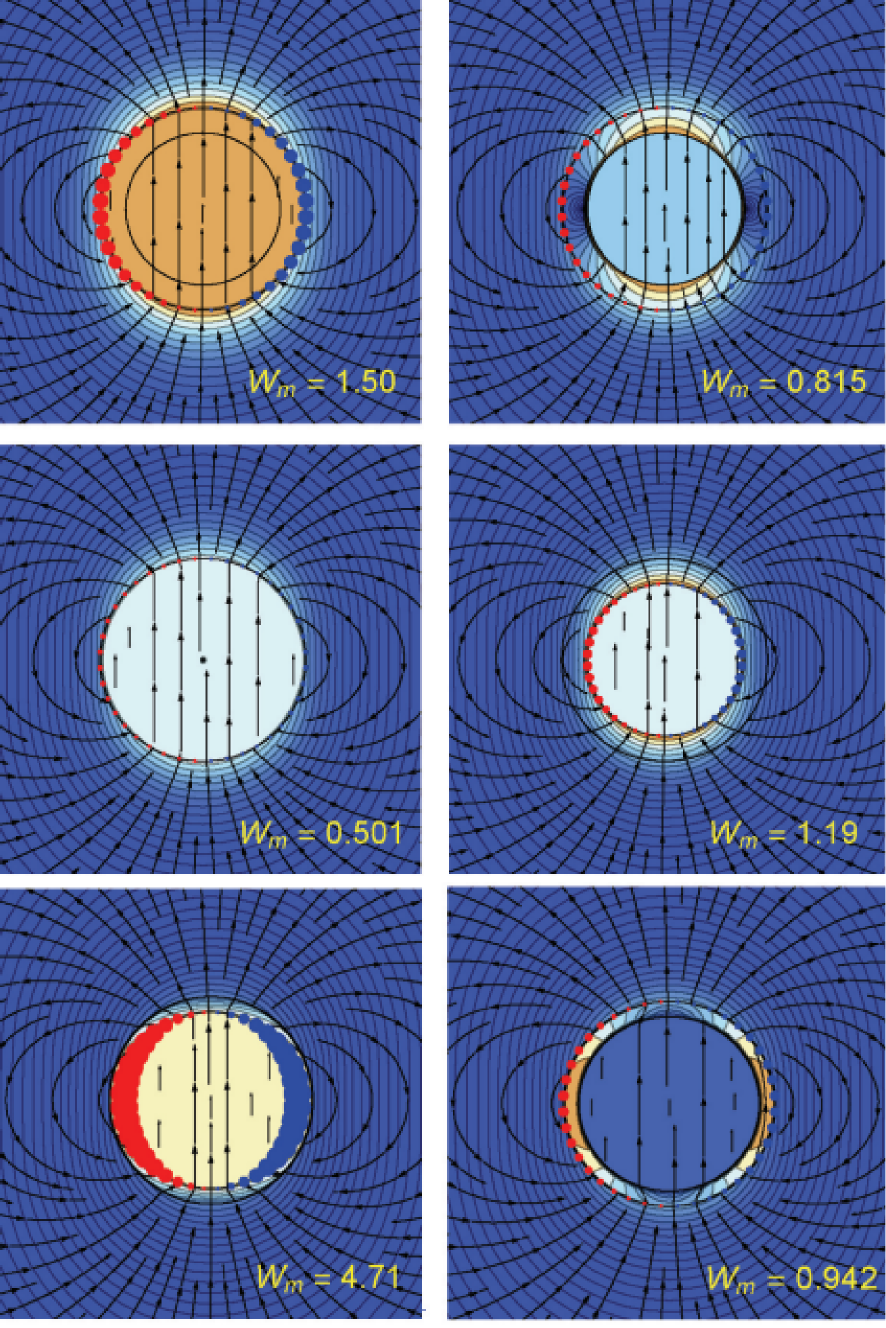}
\caption{Six different configurations of currents and paramagnetic layers producing the same magnetic dipole moment.
 Left: top: empty spherical coil, Left: middle: paramagnetic sphere with $\mu$=1000 filling spherical coil. Left: bottom: paramagnetic shell  ($0.85r_o<r<r_o$) with $\mu$=5  magnetized by a coil on the inner surface,  Right: top: paramagnetic sphere with radius 0.75$\,r_o$ and $\mu=10^3$ inside spherical coil with radius $r_o$,  Right:middle:  the same paramagnetic sphere as above but driven by current on the surface. Right: bottom: paramagnetic shell as to the left magnetized by a coil on the outer surface.  Driving current is represented by blue (for incoming) and red (for outgoing) dots with diameter proportional to current. Black streamlines follow magnetic field lines, while the grayish logarithmically scaled contour lines and the blue to brown color shading plots indicate the distribution  of $B^2$.}
\label{SL1a}
\end{figure}

A magnetic dipole can be generated in many different ways. Let us consider a spherical magnet made of paramagnetic material that is magnetized by a current flowing on a spherical surface arranged so that it produces only a dipole. Fig.~\ref{SL1a} shows six configurations of magnetizing current in a spherical magnet producing the same external magnetic field but requiring different energy input. It can be shown quite generally that the energetically least demanding way to produce a given  magnetic moment is to run the magnetizing current only on the surface of the magnet, distributed in such a way that it produces a homogeneous magnetic field inside. In this case the magnetic energy required to generate a magnetic moment $\vec {\cal M}=\int \vec M\, dV$ becomes $W_M=\mu_0\frac{{\cal M}^2}{4V}(1+\frac{2}{\mu})$, where the first term is the contribution of the surface integral  and the second of the volume integral in (\ref{magnEn}). 

A permanent magnet differs from the current driven magnet by the fact that magnetization resulting from  spontaneous alignment of elementary magnetic dipoles along a common direction,  is the only source of magnetism. Therefore, in this case $\mu\rightarrow\infty$ and the volume part of magnetic action vanishes.   
We therefore consider the magnetic action to be expressed only by the surface integral (\ref{MagSurfInt}), an integral over the compact \textbf{magnetic surface} that  encloses the volume of complete magnetization. This circumvents the classic problem of designing a classical dynamo mechanism  (\cite{1965MNRAS.131..105M}) that produces the observed magnetic field. We propose instead that the "dynamo" most likely works through the most efficient mechanism possible to produce the observed field.

In taking the limit $\mu\to\infty$ we do not assume the stellar material to be a
ferromagnet in the usual sense.  Rather, this limit represents the effective
situation in which the intrinsic magnetization of degenerate matter is large
compared to the auxiliary field $H$, so that $B\simeq\mu_0 M$ and the volume
term $\mu_0 H(H+M)$ becomes negligible.  The internal magnetic energy is then
contained in the free energy of the magnetized electron gas, while the work
associated with the external dipole field arises solely from the surface
integral.  This limit serves as a convenient mathematical device for isolating
the minimal action required to generate the observable magnetic dipole.

\subsection{Magnetic interaction energy}
\label{SSec3_2}
The above discussion shows that the magnetic contribution to the action can be
reduced to a surface term once the internal magnetization of matter is treated
as part of its equilibrium free energy.  In this macroscopic description the
volume term involving $H(H+M)$ vanishes in the effective limit $H \ll M$, which
characterizes a medium whose magnetization is intrinsic rather than maintained
by driven currents.  To justify this assumption on microscopic grounds, we now
turn to the quantum mechanical properties of degenerate matter.  Following
Chandrasekhar’s approach to the electron gas, we show that a dense
electron--ion plasma naturally develops a magnetized ground state through the
combined effects of Landau quantization, charge polarization, and the mass
asymmetry between electrons and ions.  This provides the internal mechanism
responsible for the spontaneous magnetization that underlies the surface
representation of magnetic action derived in the previous subsection.

Matter interacts with magnetic field through elementary magnetic moments of its constituent electrons ($\mu_B$), protons ($\mu_N$), neutrons ($\mu_N$). Let us consider the equation of state of  cold white dwarf matter - degenerate matter - under given pressure and immersed in a given magnetic field. 
  
In the absence of magnetic field, one starts by describing electrons in a cubic box with a wave function, which is an antisymmetric linear combination of single particle wave functions satisfying the Schr\"{o}dinger equation $-\frac{\hbar^2}{2m_e}\Delta\Psi=E\,\Psi$.  The single particle wavefunctions in a box with sides $L$ are $\psi_{pqrs}\propto \sin(\frac{p \pi}{L}x)\sin(\frac{q \pi}{L}y)\sin(\frac{r \pi}{L}z)\vert s\rangle $ with $p$, $q$, $r$  integers and the spin quantum number $s$ takes on the values $\pm\frac{1}{2}$.  Energy eigenvalues of these states are: $E_{pqns}=\frac{\hbar}{2m_e}\left(\frac{\pi}{L}\right)^2\left(p^2+q^2+n^2\right)$. The ground state of electron gas in this box is the state in which all the energy states up to the Fermi energy ($E_F$) are filled, i.e. if  $E_{pqns}\leq E_F$ for any combination of $p$, $q$, $r$ and $s$, then this state is occupied, otherwise it is not. The main results of this analysis are the relations between the Fermi energy and electron number density and  the relation between degenerate gas kinetic energy density and Fermi energy and degenerate gas pressure (\cite{2006asco.book.....H}):
\begin{align}
n_e &=\frac{1}{3\pi^2\hbar^3}\left(2m_e E_F\right)^{3/2}\, , \label{DegenDens}\\
w_e &=\frac{3}{5}n_eE_F=\frac{3}{5}\frac{(3\pi^2)^{2/3}\hbar^2}{2m_e}n_e^{5/3}\, ,\label{DegenEAll}\\
p_g &=\frac{(3\pi^2)^{2/3}\hbar^2}{5m_e}\,n_e^{5/3}\, .\label{PdegGas}
\end{align}

\begin{figure}[h]
\includegraphics[width=9 cm]{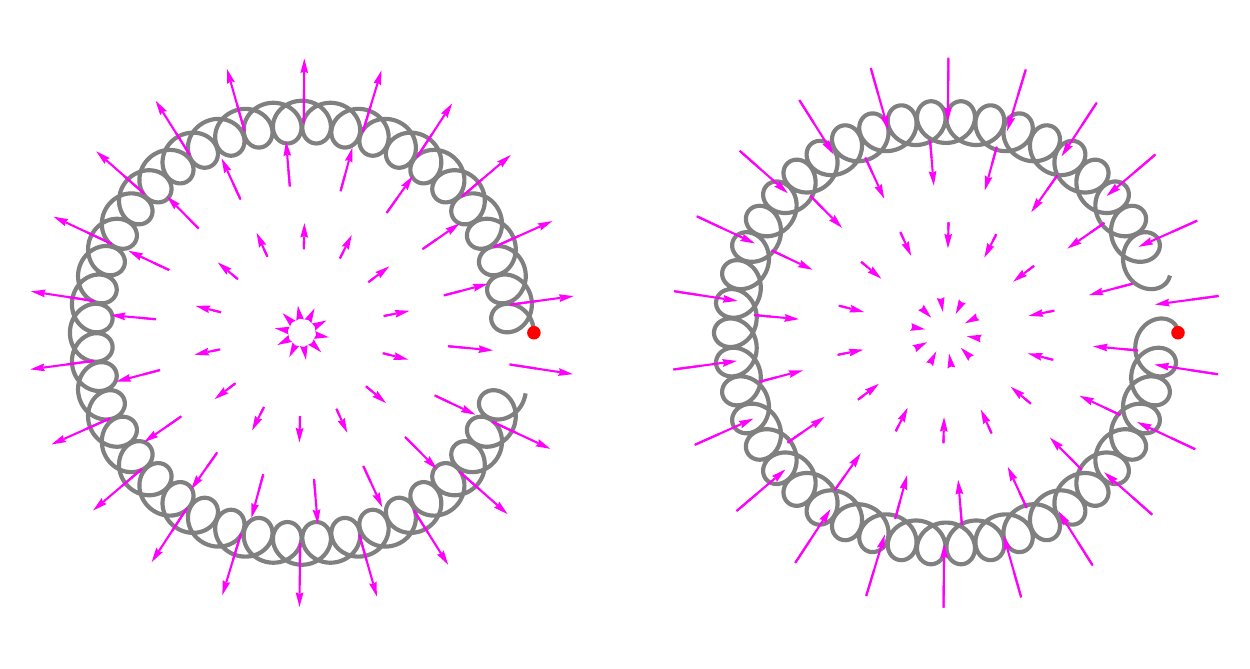}
\caption{Classical orbit corresponding to dynamical solution of $ H_{class}$ for $\varepsilon>0$ (left) and $\varepsilon<0$ (right). Orbit starting point is denoted by a red dot. Arrows in magenta depict electric field (\ref{EFieldB}). }
\label{KlasOrb}
\end{figure}

To extend this treatment to the presence of magnetic field, one must include the magnetic interaction in the Schr\"{o}dinger Hamiltonian, as well as the Weiss field, which takes account of exchange interaction (\cite{1976itss.book.....K}).  Thus, the single particle Hamiltonian must be of the form ${\cal H}=-\frac{1}{2m_e}\left(-i\hbar \nabla-e\vec A\right)^2+\frac{e}{m_e}{\bf B}\cdot {\bf \sigma}+{\cal H}_W$, where ${\cal H}_W$ represents the exchange interaction via Weiss field. In the absence of exchange interaction electron gas is diamagnetic, since electrons gyrate in magnetic field about a fixed axis in such  way as to make a current loop opposing external magnetic field. To contribute to magnetic field, an electron must close a loop in the opposite direction. This is made possible in the presence of a radial electric field, as in  Fig.\ref{KlasOrb}, which pulls the electron toward the  radial center.  

The classical Hamiltonian describing such a scenario can be written as :
\begin{equation}
{\tilde H}_{class}=\frac{1}{2m_e}\left(\vec p-e\vec A\right)^2+ m_e \tilde{\varepsilon}\, {\tilde\omega}_c^2(x^2+y^2)\, ,
\label{HamClass}
\end{equation} 
where $\vec A=\lbrace-\frac{m_e{\tilde\omega}_c}{2\,e} y,\frac{m_e{\tilde\omega}_c}{2\,e} x,0\rbrace$, $\omega_c=\frac{e\,B}{m_e}$ and $\sqrt{\tilde\varepsilon}\,{\tilde\omega}_c$ is angular velocity of gyration center circulating in the direction opposite to cyclotron gyration. The last term of the above Hamiltonian is understood to represent the (average) exchange interaction between the electrons and ions and is represented by the electric polarization:
\begin{equation}
\vec E_r=2\,{\tilde\varepsilon} \frac{m_e}{e}{\tilde\omega}_c^2\vec R\, ,
\label{EFieldB}
\end{equation} 
where $\vert \vec R\vert=\sqrt{x^2+y^2}$.
We  show at the end of this section that such electric field may reasonably be expected as a result of polarization between the cool gas of heavy positively charged ions and the degenerate gas of light electrons whoose degenerate pressure must be counterbalanced by electrostatic interaction between the two gases.

In order to quantize the above Hamiltonian, it is convenient to first express the classical Hamiltonian with another set of dynamic variables 
{\footnotesize \begin{eqnarray*}
 &\alpha=\frac{p_x-i\,p_y-\frac{i}{2}m_e\omega_c(x-i\,y)}{\sqrt{2\,m\,\tilde h\,\omega_c}}\, , &\alpha^\star=\frac{p_x+i\,p_y+\frac{i}{2}m_e\omega_c(x+i\,y)}{\sqrt{2\,m_e\,\tilde h\,\omega_c}}\, ,\nonumber \\&\beta=\frac{p_x+i\,p_y-\frac{i}{2}m_e\omega_c(x+i\,y)}{\sqrt{2\,m_e\,\tilde h\,\omega_c}}\,  ,&\beta^\star=\frac{p_x-i\,p_y+\frac{i}{2}m_e\omega_c(x-i\,y)}{\sqrt{2\,m\,\tilde h\,\omega_c}}\, ,
\end{eqnarray*}}
which belong to algebra of Poisson brackets analogous to the algebra of creation and annihilation operators in quantum mechanics\footnote{In the classical treatment ${\tilde h}$ is as  arbitrary constant}.
With these operators the Hamiltonian:
\begin{equation}
H_{class}=\hbar \omega_c\alpha\, \alpha^\star+2 \varepsilon\,\hbar \omega_c \left(\alpha\, \alpha^\star+\beta\, \beta^\star\right)+\frac{p_z^2}{2m_e} 
\label{ClassHami}
\end{equation}
is the same as (\ref{HamClass}) if $\tilde h=\frac{\hbar}{1+4\varepsilon}$, $ {\tilde\varepsilon}=\varepsilon (1+2\varepsilon)$ and ${\tilde\omega}_c=\frac{\omega_c}{1+4\varepsilon}$. 

The $z$ component of angular momentum $l_z=m(x\,\dot y-y\,\dot x)$, expressed with $\alpha$, $\beta$, becomes:
 \begin{equation}
 l_z=\hbar\left(-2(1+2\varepsilon)\alpha\,\alpha^\star+4\,\varepsilon\,\beta\,\beta^\star+\left(\alpha\,\beta+\alpha^\star\beta^\star\right)\right)\, .
 \label{AngMomClass}
 \end{equation}
The quantum Hamiltonian and angular momentum operator are now readily obtained by replacing dynamic variables $\alpha$, $\alpha^\star$, $\beta$, $\beta^\star$ with annihilation and creation operators $a$, $a^\star$, $b$, $b^\star$ and adding the spin  interaction  ($\sigma_z$)  with magnetic field\footnote{In replacing, we symmetrize the products $\alpha\, \alpha^\star\rightarrow \frac{1}{2}(a\,a^\star+a^\star\,a)\rightarrow a^\star\,a+\frac{1}{2}$,  taking into account the commutation relations for $a$ and $b$ operators.}. 
\begin{equation}
{\cal H}=\hbar\omega_c\left(a^\star a+\frac{1}{2}+2\varepsilon \left(b^\star b+a^\star a+1\right)+\sigma_z\right)-\frac{\hbar^2}{2m_e}\frac{\partial^2}{\partial z^2}\, .
\label{QuantHam}
\end{equation}
We consider  the degenerate electron gas to be constrained in a cylinder of hight $L$ and radius $R_0$. The eigenfunctions of (\ref{QuantHam}) separate in cylindrical coordinates into products $ \vert n\,s\,k_z\,\sigma_z\rangle\,\propto\,\sin  (k_z z)\vert\sigma_z\rangle \vert n,s\rangle$. The function $\sin (k_z z)$ ensures that the wave function vanishes at the bottom ($z=0$) and at the top ($z=L$) of cylindrical enclosure if $k_z=\frac{\pi}{L}q$ for $q$ an integer. The radial part of single electron wave function $\vert n,s\rangle$  can be generated by repeated action of  operators $a^+$ and $b^+$ on the vacuum state $\vert 0\rangle$: 
\begin{equation}
\vert n\,s\rangle=e^{i\,k_z z}(n!s!)^{-1/2}a^{+^n}b^{+^s}\vert 0\rangle\, .
\end{equation}
The quantum number $n$ can be thought of as quantizing the kinetic energy of the electron gyrating about a fixed axis with angular frequency $\omega_c$, while the quantum number $s$ can be thought of as quantizing the radial position of the axis of circular motion, which in this case moves (with angular frequency $\sqrt{\varepsilon}\, \omega_c$) on a circular orbit with respect to the center of rotation of  degenerate gas. The radial position $R_r$ of the axis of gyration corresponding to eigenfunction  $\vert n\,s\,k_z\,\sigma_z\rangle$ can be defined by: 
\begin{align}
R_r^2&=\langle n\,s\,k_z\,\sigma\vert (x^2+y^2)\vert n\,s\,k_z\,\sigma\rangle\nonumber\\&=\frac{\hbar}{2m_e \omega_c}\langle n\,s\,k_z\,\sigma\vert 4(a^\star-b)(a-b^\star)\vert n\,s\,k_z\,\sigma\rangle\nonumber\\&=\frac{\hbar}{2m_e \omega_c}4(1+n+s)
\end{align}
and the effective width of the wave  is the gyration radius which is (for small $\varepsilon$):
$\delta R_r\sim \sqrt{\frac{\hbar}{2m_e \omega_c}n}$. If the radius $R_0$ of the cylindrical enclosure is sufficiently large with respect to $\delta R_r$, then the presence of an electron inside the enclosure can be quite accurately confirmed if $R_r\leq R_0$ and denied if  $R_r> R_0$. 
\begin{figure}[h]
\includegraphics[width=8cm]{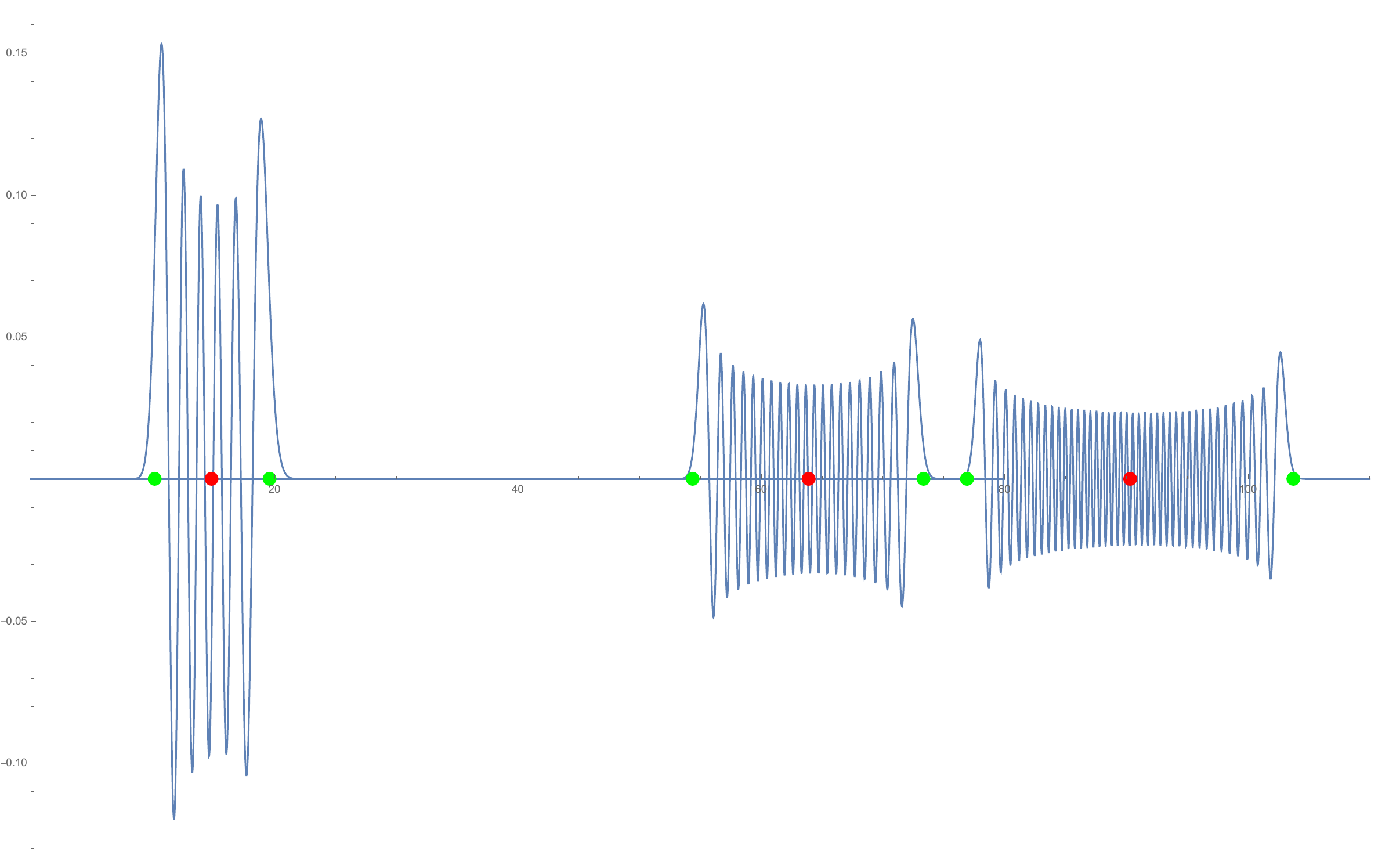}
\caption{Wave functions $\vert n,s\rangle $ for (n=10, s=100), (n=40, s=2000) and (n=80, s=4000) . Red points are at the center ${\tilde R}_c=\sqrt{2(n+s-1)}$ and green points mark the width at ${\tilde R}_c\pm \frac{3}{2}\sqrt{n}$\,. }
\label{LandW}
\end{figure}
This condition, together with the requirement $R_0\gg \delta R_r$, replaces the commonly used boundary condition $\Psi=0$ at boundary surface.

The validity of this approach is supported by the identity:
\begin{equation}
\displaystyle \sum_{s=0}^\infty \langle n,s,k_z,\sigma_z\vert \delta^2(R-R_0)\delta(z-z_0)\vert n,s,k_z,\sigma_z\rangle=\frac{m_e\omega_c}{2\pi \hbar L}\, ,
\label{DensityStates}
\end{equation}
which guarantees the completeness of Hilbert space, but also assures that the incoherent density matrix $\displaystyle \sum_{s=0}^\infty \vert n,s,k_z,\sigma_z\rangle \langle n,s,k_z,\sigma_z\vert$ produces a homogeneous density of electrons in the slab $0<z<L$ \citep{1981Ap&SS..77..299C}.

To calculate the number of particles, the energy and angular momentum of electrons in the ground state inside the cylindrical box, we must fill all the levels below the Fermi energy according to zero temperature Fermi-Dirac distribution, noting that the single state energy level must be gauged with respect to ground level of chemical potential.

The eigenfunctions and eigenvalues of the single state Hamiltonian are:
\begin{align}
{\cal H}&\vert n\,s\,k_z\,\sigma_z\rangle=\\&\left(\hbar \omega_c\left(n+\frac{1}{2}+2\,\varepsilon\left(n+ s+1\right)+\sigma_z\right)+\frac{\hbar^2}{2m_e L^2}q^2\pi^2\right)\vert n\,s\,k_z\,\sigma_z\rangle\, .
\nonumber
\end{align}
It is convenient to introduce the quantized dimensionless radial coordinate ${\cal R}^2=\frac{m_e\omega_c}{4\hbar}R^2=(n+s+1)$ and the scale factor $\zeta^2=\left(\frac{\pi}{L}\right)^2\frac{\hbar}{2m_e\omega_c}$, which casts the eigenvalues of the Hamiltonian in the form:
\begin{equation}
E_{n,{\cal R},q,\sigma_z}=\hbar\,\omega_c\left(n+\frac{1}{2}+\varepsilon {\cal R}^2+\zeta^2 q^2+ \sigma_z\right)\, .
\end{equation}

In order to determine the ground level of chemical potential, the quantum Hamiltonian must be split into the kinetic and potential part. We do this in analogy with the equivalent procedure in classical mechanics by first defining the quantum equivalent of velocity components  as:
{\footnotesize \begin{align}
{\hat v}_x&=-\frac{i}{\hbar}\left(x{\cal H}-{\cal H}x\right)=\sqrt{\frac{\hbar \omega_c}{2m_e}}\left((1+2\varepsilon)(a+a^\star)+2\varepsilon(b+b^\star)\right)\, ,
 \nonumber \\
{\hat v}_y&=-\frac{i}{\hbar}\left(y{\cal H}-{\cal H}y\right)=i\sqrt{\frac{\hbar \omega_c}{2m_e}}\left((1+2\varepsilon)(a-a^\star)+2\varepsilon(b-b^\star)\right)\, ,\nonumber \\
{\hat v}_z&=-i\,\hbar\frac{\partial}{\partial z}\, .
\label{HitrostY}
\end{align}}
With these the quantum Hamiltonian (\ref{QuantHam}) can be written as:
{\footnotesize \begin{align}
{\cal H}&=\frac{m}{1+4\varepsilon}\left(\frac{1}{2}({\hat v}_x^2+{\hat v}_y^2+(1+4\varepsilon){\hat v}_z^2)+\varepsilon(1+2\varepsilon)\omega_c^2({\hat x}^2+{\hat y}^2)\right)+\hbar \omega_c \,\sigma_z\nonumber\\
&=\left[\hbar\omega_c\left(a^\star a+\frac{1}{2}+\frac{4\varepsilon^2}{1+4\varepsilon}\left(a^\star a+b^\star b+1\right) \right .\right.\nonumber\\&\left .\left. \qquad\qquad\qquad+2\varepsilon\frac{1+2\varepsilon}{1+4\varepsilon}\left(a\,b+a^\star b^\star\right)\right)+\frac{1}{2}m\,{\hat v}_z^2\right]\nonumber\\\qquad
&\qquad\qquad+\left[\varepsilon\frac{1+2\varepsilon}{1+4\varepsilon}m\,\omega_c^2({\hat x}^2+{\hat y}^2)\right]+\hbar \omega_c \,(\sigma_z-\frac{1}{2})\, .
\end{align}}
The expression in the first square brackets represents the kinetic energy, and the one in the second the potential energy due to electric polarization. 
In the same manner we derive the operator corresponding to the $z$ component of angular momentum:
\begin{align}
{\hat l}_z&=m\left(x\,{\hat v}_y-y\,{\hat v}_x\right)=2\hbar\left((1+2\varepsilon)a^\star a-2\varepsilon\, b^\star b-\frac{1}{2}(a\,b+a^\star b^\star)+\frac{1}{2}\right)\, .
\label{QuantLZ}
\end{align}
The single state energy with respect to the ground level of chemical potential is the kinetic part of energy eigenvalue  $\langle n\,s\,k_z\,\sigma_z\vert {\cal H}_{kin}\vert\ n\,s\,k_z\,\sigma_z\rangle=\hbar \omega_c \left(n+\frac{1}{2}+\zeta^2q^2+\sigma_z\right)$. 

The number of electrons in the cylinder  can now be expressed with the sum:
\begin{align}
N&=\displaystyle\sum_{{\cal R}^2=0}^{\frac{m_e\omega_c}{2\hbar}R_0^2}\displaystyle\sum_{n,q,\sigma_z}
 \begin{cases}
    1, \quad\text{if } \frac{E_F}{\hbar\omega_c}\geq n+\frac{1}{2}+\zeta^2q^2+\sigma_z \label{ElNumDen} \\               
    0,\quad\text{otherwise}
\end{cases}\nonumber\\
&=\frac{m_e\omega_c\,R_0^2}{3\hbar \zeta}\left(\left(\frac{E_F}{\hbar\,\omega_c}\right)^{3/2}+\left(\frac{E_F}{\hbar\,\omega_c}-1\right)^{3/2}-2\zeta^3\right)\nonumber\\
&= \frac{1}{3\pi^2\hbar^3}\left(2m_e E_F\right)^{3/2}\left(1-\frac{3}{4}\frac{\hbar\,\omega_c}{E_F}+\dots\right)\left(\pi R_0^2\,L\right)\,+\,\frac{\pi^2}{3\sqrt{2}}\frac{R_0^2}{L^2}\, ,
\end{align}
which agrees with (\ref{DegenDens}) for $\omega_c\rightarrow 0$. In a similar manner we calculate the energy  (using notation ${\cal R}_0=\sqrt{\frac{m_e \omega_c}{2\hbar}}R_0$): 
\begin{align}
W_e&=\displaystyle\sum_{{\cal R}^2=0}^{\frac{m_e\omega_c}{2\hbar}R_0^2}\displaystyle\sum_{n,q,\sigma_z}
 \begin{cases}
    E_{n,{\cal R},q,\sigma_z}, \quad\text{if } \frac{E_F}{\hbar\omega_c}\geq n+\frac{1}{2}+\zeta^2q^2+\sigma_z  \\               
    0,\quad\text{otherwise}
\end{cases}\nonumber \\
&=\left(\pi\,R_0^2L\right)\frac{1}{15\sqrt{2}\pi^2}\left(\frac{m_e\omega_c}{\hbar}\right)^{3/2}\hbar \omega_c\left[\left(6\frac{E_F}{\hbar \omega_c}+5\varepsilon \,{\cal R}_0^2\right)\left(\frac{E_F}{\hbar \omega_c}\right)^{3/2}\right.\nonumber\\& \left. 
\,+
\,\left(6\frac{E_F}{\hbar \omega_c}+5\,\varepsilon\,{\cal R}_0^2+4\right)\left(\frac{E_F}{\hbar \omega_c}-1\right)^{3/2}\right]\nonumber\\
&=\left(\pi\,R_0^2L\right)\frac{3}{5}E_F\frac{1}{3\pi^2\,\hbar^3}\left(2m_e\,E_F\right)^{3/2}\nonumber\\& \left[1+\frac{5}{12}(2\,\varepsilon\,{\cal R}_0^2-1)\frac{\hbar\,\omega_c}{E_F}-\frac{5}{16}(2\,\varepsilon\,{\cal R}_0^2-1)\left(\frac{\hbar\,\omega_c}{E_F}\right)^2+\dots\right]\, ,
\label{DegGasEnergy}
\end{align}
which becomes (\ref{DegenEAll}) if $\omega_c\rightarrow 0$ and ${\varepsilon}\rightarrow 0$.

The calculation of angular momentum of electron gas requires a closer scrutiny, since the eigenfunctions of energy $\vert n, s,q,\sigma_z\rangle$ are not eigenfunctions of the quantized version of angular momentum operator (\ref{QuantLZ}). The significance of this fact can be understood by the analysis  of the classical Hamiltonian system ${\tilde H}_{class}$, which leads to the solution: $\alpha=a_n\,e^{i\omega_1 t}$, $\alpha^\star=a^\star_n\,e^{-i\omega_1 t}$, $\beta=b_s\,e^{i\omega_2 t}$ and $\beta^\star=b^\star_s\,e^{-i\omega_2 t}$, where $\omega_1=\frac{1}{2}\left(1+\sqrt{1+8\varepsilon}\right)\omega_c$ and $\omega_2=\frac{1}{2}\left(1-\sqrt{1+8\varepsilon}\right)\omega_c$, so that $l_z=2\hbar(-\vert\alpha_n\vert^2+{\tilde\varepsilon}\vert\beta_s\vert^2+(\frac{1}{2}-{\tilde{\varepsilon}})(\alpha_n\beta_s\,e^{i\,\omega_ct}+\alpha^\star_n\beta^\star_s\,e^{-i\,\omega_ct}))$. Thus, the angular momentum of an electron oscillates about the average value  $l_z=2\hbar(-\vert\alpha_n\vert^2+{\tilde\varepsilon}\vert\beta_s\vert^2)$ with the frequency $\omega_c$. This means that a single electron exchanges angular momentum with electromagnetic field\footnote{This mechanism is used in magnetrons to generate electromagnetic waves}. As a consequence, electrons in a magnetized degenerate gas exchange angular momentum among themselves via this interaction. A rigorous treatment of magnetized degenerate gas may require quantizing electromagnetic field as well. However, since the interaction of electrons with magnetic field in a magnet does not produce real photons, it seems justified to assume that angular momentum is exchanged between electrons with virtual photons only, representing a static electric and magnetic field. Thus, we can assign to each electron the average angular momentum 
$\langle l_z\rangle=\hbar\langle n,s,q,\sigma_z\vert \left(-2a\,a^\star +4{\tilde\varepsilon}b\,b^\star+\sigma_z\right)\vert n,s,q,\sigma_z\rangle= \hbar \left(-2(1+4\varepsilon)(n+\frac{1}{2})+2\varepsilon\,{\cal R}^2+\sigma_z\right)$. This allows us to write down the angular momentum density as\footnote{One should keep in mind that the limit $\omega_c\rightarrow \infty$ does not exist, since in this case  the density matrix (\ref{DensityStates}) becomes singular, therefore, the boundary condition  $\delta R\ll R_0$ can not be met. However, for all $\omega_ c\neq 0$ the angular momentum density has a definite value for magnets that are much, much larger than the cyclotron radius $\frac{2\hbar}{m_e\omega_c}$.}:
\footnotesize{
\begin{align}
\lambda_z&=\frac{1}{\pi\,R_0^2L}\displaystyle\sum_{{\cal R}^2=0}^{\frac{m_e\omega_c}{2\hbar}R_0^2}\displaystyle\sum_{n,q,\sigma_z}
\begin{cases}
 \langle l_z\rangle ,\quad\text{if } \frac{E_F}{\hbar\omega_c}\geq n+\frac{1}{2}+\zeta^2q^2+\sigma_z  \\               
   0,\quad\text{otherwise}
\end{cases}\nonumber \\
&=\left(\frac{m_e\omega_c}{\hbar}\right)^{3/2}\left[ \frac{\sqrt{2}}{3\pi^2}\,\varepsilon\,{\cal R}^2\left(\left(\frac{E_F}{\hbar \omega_c}-1\right)^{3/2}+\left(\frac{E_F}{\hbar \omega_c}\right)^{3/2}\right)\right.\nonumber\\
& -\frac{1}{15\sqrt{2}\pi^2}\left(\frac{E_F}{\hbar \omega_c}-1\right)^{3/2}\left(8\frac{E_F}{\hbar \omega_c}-3\right)\nonumber \\&-\frac{1}{15\sqrt{2}\pi^2}\left(\frac{E_F}{\hbar \omega_c}\right)^{3/2}\left(8\frac{E_F}{\hbar \omega_c}+15\right)
\nonumber\\
&\left.+\frac{4\sqrt{2}}{15\pi^2}\varepsilon \left(\left(\frac{E_F}{\hbar \omega_c}-1\right)^{3/2}\left(4\frac{E_F}{\hbar \omega_c}+1\right)+\left(\frac{E_F}{\hbar \omega_c}\right)^{3/2}\left(4\frac{E_F}{\hbar \omega_c}+5\right)\right)\right]\, .
\nonumber
\end{align}}

\newpage
Taking into account that an electron in a state with angular momentum ${\mathrm l}_z=\hbar \, l$ behaves as a magnetic dipole $\mu_z=\mu_B\, l$ (where $\mu_B$ is Bohr magneton), one finds  that $M_z=\mu_B \lambda_z/\hbar$ is the magnetization generated by the degenerate gas. If $\mu_0 M_z=B$, then the only source of magnetic field is the degenerate gas itself. This condition is met if 
\begin{equation}
\varepsilon=\frac{\frac{1}{2}\left(\frac{E_F}{\hbar \omega_c}-1\right)^{3/2}\left(8\frac{E_F}{\hbar \omega_c}-3\right)+\left(\frac{E_F}{\hbar \omega_c}\right)^{3/2}\left(8\frac{E_F}{\hbar\omega_c}+15\right)+\frac{30\pi^2}{\sqrt{2\tau_0\omega_c}}}{\left(\frac{E_F}{\hbar \omega_c}-1\right)^{3/2}\left(5{\cal R}^2-16\frac{E_F}{\hbar \omega_c}-4\right)+\left(\frac{E_F}{\hbar \omega_c}\right)^{3/2}\left(5{\cal R}^2-16\frac{E_F}{\hbar \omega_c}-20\right)}\, ,
\label{EpsilonB}
\end{equation}
where $\tau_0=\frac{4m_e}{\hbar}\pi^2\,r_{cl}^2$, $r_{cl}$ is classical electron radius. The energy  of degenerate electron gas, which supports its own magnetic field can thus be expressed with (\ref{DegGasEnergy}) by replacing $\varepsilon $ with (\ref{EpsilonB}) and taking the limit ${\cal R}^2\gg \frac{E_F}{\hbar\omega}$. After some algebra we obtain ($w_e=\frac{W_e}{\pi R_0^2L}$): 

\begin{align}
w_e =&E_F\frac{1}{24\pi^2\hbar^3}\left(2 m_e E_F\right)^{3/2}\left[\left(1-\frac{\hbar \omega_c}{E_F}\right)^{3/2}\left(4+\frac{\hbar \omega_c}{E_F}\right)\right. \nonumber\\+&\left(4+3\frac{\hbar \omega_c}{E_F}\right) \left. +\frac{6\pi^2}{\sqrt{2\tau_0\omega_c}}\left(\frac{\hbar \omega_c}{E_F}\right)^{5/2}+\frac{E_F}{m_e c^2}{\cal O}\left(\frac{8 c}{5 R_0\omega_c}\right)^2\right]\nonumber\\
\rightarrow & E_F\left(\frac{1}{3\pi^2\hbar^3}\left(2m_e E_F\right)^{3/2}\right)\left[1-\frac{1}{4}\frac{\hbar \omega_c}{E_F}
 \right. \nonumber\\&\left.
+\frac{3\pi^2}{\sqrt{\omega_c\tau_0}}\left(\frac{\hbar \omega_c}{2E_F}\right)^{5/2}+\dots+\frac{E_F}{m_e c^2}{\cal O}\left(\frac{8 c}{5 R_0\omega_c}\right)^2\right]\, .
\label{EnergDensPlus}
\end{align}

In a similar way we express the  components of electron gas stress tensor by quantizing the classical expression  $T_{i\,j}=\displaystyle\sum_{k=1}^{n_e}m_e\,(\underset{k}{v})_i(\underset{k}{v})_j$, which is a sum over particles in a unit volume. The components   $T_{ij}$ are expressed in the same form as the energy (\ref{DegGasEnergy}) if one replaces
 $E_{n,{\cal R},q,\sigma_z}$ with expectation values  $\langle n,s,k_z,\sigma_z \vert m_e\,{\hat v}_i {\hat v}_j \vert n,s,k_z,\sigma_z\rangle$, where electron velocity components are expressed with eqs.(\ref{HitrostY}), so that:
\begin{align}
 m_e\left({\hat v}_x^2+{\hat v}_y^2\right)&=\hbar\omega_c\left(\left(1+\frac{4\varepsilon^2}{1+4\varepsilon}\right)\left(a^\star a+\frac{1}{2}\right)+\frac{4\varepsilon^2}{1+4\varepsilon}\left(b^\star b+\frac{1}{2}\right)\right)\nonumber\\&+2\varepsilon \frac{1+2\varepsilon}{1+4\varepsilon}\left(a\,b+a^\star b^\star\right)\, ,
\nonumber
\end{align} 
\begin{equation}
 m_e{\hat v}_z^2=-\frac{1}{2}\hbar^2\frac{\partial^2}{\partial z^2}\, ,\nonumber
\end{equation} 
\begin{align} 
m_e {\hat v}_x{\hat v}_y&=\frac{i}{2}\hbar \omega_c\left(\left(1+2\varepsilon\right)^2\left(a\,a-a^\star a^\star\right)\right.\nonumber \\ &\left.+4\varepsilon \left(1+2\varepsilon\right)\left(b^\star a-b\,a^\star\right)+4\varepsilon^2(b^\star b^\star-b\,b)\right)\, .\nonumber
\end{align}
 The expectation values of these operators lead to the following stress tensor components:
\begin{align}
T_{x,x}&=T_{y,y}=\frac{2}{15\pi^2}\left(\frac{2m_e}{\hbar^2}\right)^{3/2}E_f^{5/2}\left(1+\frac{5}{64}\left(\frac{\hbar \omega_c}{E_F}\right)^3+\dots\right)\, ,
\label{PdegGX}\\
T_{z,z}&=\frac{2}{15\pi^2}\left(\frac{2m_e}{\hbar^2}\right)^{3/2}E_f^{5/2}\left(1-\frac{5}{4}\frac{\hbar \omega_c}{E_F}+\frac{15}{16}\left(\frac{\hbar \omega_c}{E_F}\right)^2+\dots\right)\, .
\label{PdegGZ}
\end{align}
Note that equations (\ref{PdegGX}) and (\ref{PdegGZ}) reduce to  (\ref{PdegGas}) for $\omega_c\rightarrow 0$.

\begin{figure}[h]
\includegraphics[width=8.5cm]{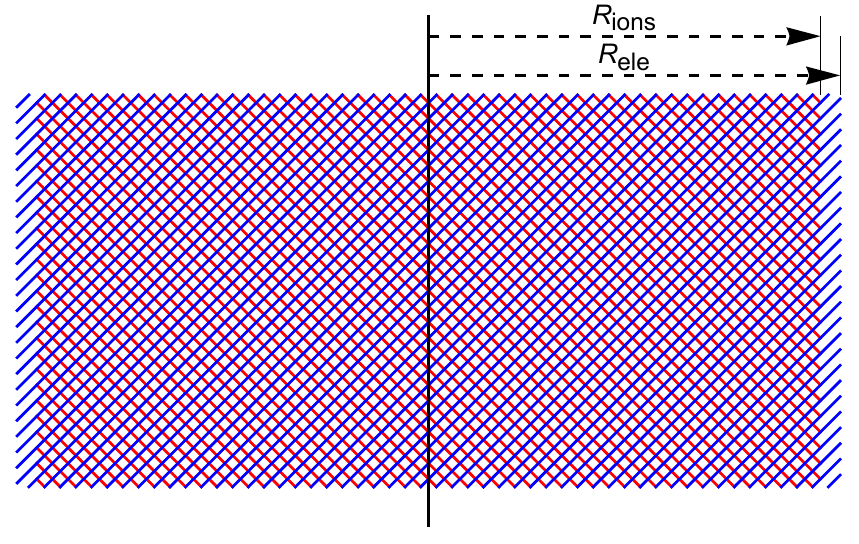}
\caption{The gas  of ions (red) occupies a smaller cylinder than the gas of electrons (blue). From $n_e R_{ele}^2-n_i R_{ion}^2=0$ it follows: $(n_e-n_i)R_{ele}^2\approx 2n_i R_{ele}\delta R$, where $
\delta R=R_{ele}-R_{ions}$. }
\label{CharImbal}
\end{figure}

The main lesson taught by inclusion of angular momentum in the interaction of electrons with magnetic field is the recognition that energy eigenstates are not eigenstates of angular momentum, since in the presence of magnetic field electrons exchange angular momentum with magnetic field as well as with the electric field of ions. Electrons, having more energy per particle than heavy ions in a degenerate gas, occupy a slightly larger volume than ions, creating a surface charge layer, as shown in  Fig.\ref{CharImbal}. This  interaction generates the static electric polarization (\ref{EFieldB}).

To estimate the thickness ($\delta R$) of the surface charge layer, note that the charge imbalance 
 $\delta \rho=e\left(n_e-n_i\right)=\varepsilon_0\nabla\cdot {\vec E}_r$. Expressing  $\tilde\varepsilon\approx\varepsilon$ in (\ref{EFieldB}) with (\ref{EpsilonB})   for very large ${\cal R}^2$ and  assuming $E_F\gg \hbar\omega_c$, we obtain $\delta\rho=\frac{16}{5}\varepsilon_0\frac{1}{{\cal R}^2}\frac{m_e\,\omega_c\,E_F}{e\,\hbar}=\frac{16}{5}\frac{\varepsilon_0\,E_F}{e \,R_{ele}^2}$. Overall charge neutrality requires $\frac{\delta \rho}{e\, n_e}=2 \frac{\delta R}{R}$, so that the thickness of the surface charge layer is $\delta R\sim \frac{ (48\lambda_c)^2}{R_{ele}}\sqrt{\frac{m_e c^2}{E_F}}$, where $\lambda_c=\frac{h}{m_e c}$. This represents a very thin layer as can be seen by expressing the radius of the sample in terms of the number of cyclotron radii ($N_c$) as $R_{ele}\sim N_c \sqrt{\frac{E_F}{m_e\omega_c^2}}$ and Fermi energy as $E_F=n_{max}\hbar \omega_c$, then $\delta R\sim\frac{33.9}{N_c n_{max}}\lambda_c$, i.e. it is only a fraction of Compton wavelength, since $N_c\gg 1$ and $n_{max}>1$.  


Finally, we must express the total energy and pressure of the magnetized degenerate electron gas as a function of electron density and magnetic field density. We first express Fermi energy by inverting eq.(\ref{ElNumDen}) in a series expansion:
\begin{align}
E_F&=\frac{(3\pi^2)^{2/3}\hbar^2}{2m_e}n_e^{2/3}\left[1 - \frac{1}{(9\pi)^{2/3}}\left(\frac{\omega_c}{\lambda_c c\,n_e^{2/3}}\right)^2\right.\label{FermiMagAll}\\&\left.- \frac{1}{(9\pi)^{4/3}}\left(\frac{\omega_c}{\lambda_c c\,n_e^{2/3}}\right)^4- \frac{7}{3(9\pi)^{6/3}}\left(\frac{\omega_c}{\lambda_c c\,n_e^{2/3}}\right)^6+\dots\right]+\frac{1}{2}\hbar\,\omega_c \,. \nonumber
\end{align}
The energy density (\ref{EnergDensPlus}) then becomes:
\begin{align}
&w_e=\nonumber \\ &\frac{(3\pi^2)^{2/3}\hbar^2}{2m_e}n_e^{5/3}+n_e\,\hbar\,\omega_c+\frac{m_e\,\omega_c^2}{4\pi\,r_{cl}}\left(1+\left(\frac{n_e r_{cl}^3}{9\pi}\right)^{1/3}\right)\nonumber \\&-\frac{m_e^3\omega_c^4}{96\pi^4\hbar^2 n_e}+\dots  \label{DegenEMag} \\
&=\frac{(3\pi^2)^{2/3}\hbar^2}{2m_e}n_e^{5/3}+\frac{B^2}{\mu_0}\left[1+\frac{2\pi}{\alpha_f^2}\frac{m_e c^2}{\mu_BB}(n_e r_{cl}^3)+\left(\frac{1}{9\pi}n_e r_{cl}^3\right)^{1/3}\right.\nonumber \\&\left.-\frac{\alpha_f^2}{6\pi^2}\frac{B^2/\mu_0}{n_e m_e c^2}\right]+\dots
\nonumber
\end{align}

and the components of pressure tensor (\ref{PdegGX}, \ref{PdegGZ} ) turn into:
\begin{align}
T_{x,x}&=T_{y,y}=\nonumber\\ &\frac{(3\pi^2)^{2/3}\hbar^2}{5\,m_e}n_e^{5/3}+\frac{1}{2}\hbar\,\omega_c\,n_e\nonumber \\& +\frac{1}{4}m_e\,\omega_c^2\left(\frac{n_e}{9\pi^4}\right)^{1/3}- \frac{m_e^3\omega_c^4}{72\pi^4\hbar^2 n_e}+\dots\label{PdegenMagX}\\
T_{z,z}&=\frac{(3\pi^2)^{2/3}\hbar^2}{5\,m_e}n_e^{5/3}+\frac{1}{4}m_e\,\omega_c^2\left(\frac{n_e}{9\pi^4}\right)^{1/3}+\nonumber\\&\qquad+\frac{m_e^2\omega_c^3}{12(3\pi^8 n_e)^{1/3}\hbar}-\frac{m_e^3\omega_c^4}{72\pi^4\hbar^2 n_e}+\dots
\label{PdegenMagZ}
\end{align}

Comparing eq.(\ref{DegenDens}) with  eq.(\ref{FermiMagAll}) and eq.(\ref{PdegGas}) with eqs.(\ref{PdegenMagX},\ref{PdegenMagZ}), one notes that in the absence of magnetic field they are the same. We note that the anisotropy of the stress components in Eqs. (\ref{PdegenMagX}) and (\ref{PdegenMagZ}) is
precisely the form expected from magnetostriction, with magnetic tension along
the field direction and enhanced transverse pressure. However,  equation (\ref{DegenEAll}) assigns the energy $\frac{3}{5}E_F$ to the average electron, while in eq.(\ref{DegenEMag}) the average energy per electron is $E_F$. This difference is clearly due to the fact that eq.(\ref{DegenEAll}) only includes the kinetic energy of electrons, while eq.(\ref{DegenEMag}) includes also the potential energy of electrons with respect to electric polarization, which must be present in the degenerate gas of charged electrons immersed in the cold gas of oppositely charged ions. This electric polarization is needed to allow electrons to occupy states with both positive and negative angular momentum. 

\begin{figure}[h]
\includegraphics[width=8.5cm]{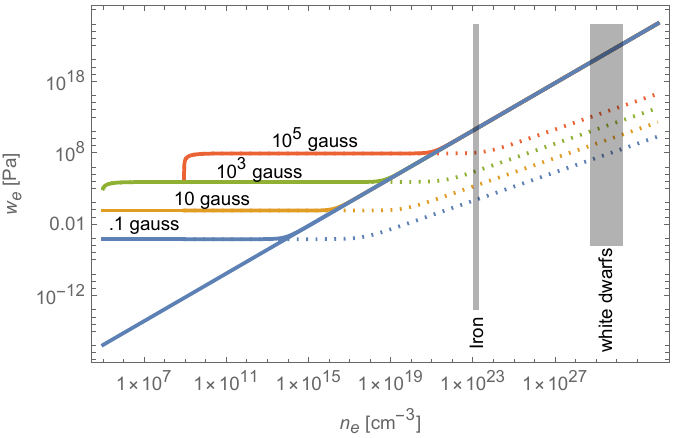}
\caption{Energy density of magnetized degenerate gas of electrons for different values of magnetic field as a function of electron density. Dotted lines represent the part due to magnetization. The gray boxes indicate conduction electron density in iron and typical average electron density in white dwarfs.}
\label{MagEnDensP}
\end{figure}

Energy density (\ref{DegenEMag}) as a function of electron number density for electron gas in equilibrium with magnetic field of different strenghts is shown in Fig.\ref{MagEnDensP}.
According to this figure one can identify three distinct parts of the curve for each magnetic field: 1) at low electron number density the last term in eq.(\ref{DegenEMag}) prevails, so that the electron gas can not support the required magnetic field; 2) at higher densities the term $\frac{m_e \omega_c}{4\pi r_{cl}}=\frac{B^2}{\mu_0}$ prevails,  making the electron gas magnetic field dominated and can not be controlled by changing its pressure; 3) only at high enough density the first term in eq.(\ref{DegenEMag}) prevails, making the degeneracy pressure the dominant force, yet in this regime the magnetic contribution ($n_e\hbar\,\omega_c$) starts increasing linearly with density (dotted lines in Fig.\ref{MagEnDensP}) and is only overtaken by degeneracy pressure because of its $5/3$ exponent. 
We note that the state of degenerate electron gas in the cores of white dwarfs is in this third regime where gravitational balance essentially dictates the local value of Fermi energy. In this case the energy density can be expressed with eq.(\ref{EnergDensPlus}), which shows that under said conditions the energy density has a minimum with respect to magnetic field ($\omega_c$) at:
\begin{equation}
B_{min}=\frac{m_e}{3\pi^2\sqrt{2}e}\sqrt{\tau_0\left(\frac{E_F}{\hbar}\right)^3}=\frac{1}{4}\mu_0\mu_B n_e\, ,
\label{MagnMin}
\end{equation}
which is by $\frac{1}{\mu_0}B^2$ below the value at $B\rightarrow 0$. We also note that any realistic stellar model has an atmospere, where electron density is too low to contribute its own magnetization.

In summary, the microscopic analysis of the magnetized degenerate electron gas
confirms the macroscopic assumptions introduced in subsection~III\,A.  The
interaction between Landau quantization, the small charge imbalance induced by
the mass asymmetry of electrons and ions, and the resulting polarization field
leads to a magnetized ground state in which $B \simeq \mu_{0} M$ and the
auxiliary field $H$ is negligible.  Thus the internal magnetic energy is
naturally incorporated into the free energy of matter, while the work required
to establish the observable magnetic dipole resides in the surface term
identified earlier.  The two approaches therefore describe the same physical
phenomenon at different levels: subsection \ref{SSec3_1} provides the minimal-action
macroscopic representation, and subsection \ref{SSec3_2} supplies the corresponding
microscopic mechanism that produces intrinsic magnetization in degenerate
stellar matter.

\section{``Lane-Emde equation'' for a  magnetic star }
\label{MagnLaEm}
In subsection (\ref{SSec3_1}) we argued that the contribution of magnetic field to the action can be expressed only by the surface integral, while in subsection (\ref{SSec3_2}) we showed that the degenerate electron gas can generate magnetization in equilibrium with electromagnetic field. The main effect of such equilibrium is the shift of internal energy of electron gas, but it  does not essentially change the equation of state (eq.\ref{PdegGas} vs. \ref{PdegenMagX},\ref{PdegenMagZ} ) except for the  pressure anisotropy. Lane-Emde equation, which follows from  the local equilibrium equations (\ref{LocEqil}), is not expected to change if magnetic field is added to action. However, at the magnetic surface local equilibrium changes, since the electromagnetic stress  is discontinous at the magnetic surface and produces additional pressure, which must be compensated by gravity. The matching conditions at the magnetic surface follow directly from the
requirement that the total stress--energy tensor of matter, gravity, and
electromagnetism has vanishing divergence in the linearized theory
\cite{MTW}.

Consider a pillbox near the surface of a star. For any box completely inside or completely outside the star the sum of forces acting on a pillbox vanish. This statement can be expressed by surface integrals of their corresponding stress energy tensors gravitational, matter\footnote{here designated with p, because its space components represent pressure} and electromagnetic as: 
$\int_S \left( {\underline{\underline {T_g}}}+ {\underline{\underline {T_p}}}+ {\underline{\underline {T_m}}}\right) \cdot d\vec S =0$.
In other words, the sum of divergencies of these tensors vanishes
\begin{equation}
{{\underset{g}{T}}^{\mu\nu}}{ ._\nu}+{{\underset{p}{T}}^{\mu\nu}}{ ._\nu}+{{\underset{m}{T}}^{\mu\nu}}{ ._\nu}\,=\,0\, .
\label{inequil}
\end{equation}
At the magnetic surface the sum of stress energy tensors can be expressed as 
$\underset{all}{T}^{\mu\nu}=\left(\underset{g}{T}^{\mu\nu}+\underset{p}{T}^{\mu\nu}+\underset{m}{T}^{\mu\nu}\right)_{ext}
\Theta(\xi-1)+\left(\underset{g}{T}^{\mu\nu}+\underset{p}{T}^{\mu\nu}+\underset{m}{T}^{\mu\nu}\right)_{int}\Theta(1-\xi)$,  
where $\Theta$ represents the Heaviside function as the function of the  radial coordinate $\xi$, such that $\xi=1$ is at the surface of the star. Since the divergence of $\underset{all}{T}\quad$ vanishes except at the surface, the divergence at the surface boils down to 
\begin{equation}
\underset{all}{T}^{\mu\nu}._\nu=\left[\left(\underset{m}{T}^{\mu\xi}+\underset{p}{T}^{\mu\xi}\right)_{ext}-\left(\underset{p}{T}^{\mu\xi}+\underset{m}{T}^{\mu\xi}\right)_{int}\right] \delta(\xi-1)\, ,\label{BdCondMR}
\end{equation} 
where we have taken into account the fact that the gravitational stress-energy tensor is continuous across the boundary.


In order to express these stress tensors, it is convenient to use the covariant formalism of linearized gravity.  The matter tensor can be considerd as the sum  of the isotropic pressure tensor, expressed with respect to arbitrary static coordinates ($x^\mu$) as $\underset{p}{T}^{\mu\nu}=\frac{\rho c^2+p}{c^2}u^\mu u^\nu+p g^{\mu \nu}$  and a weak anisotropic component described by the the anisotropic pressure induced by electromagnetic field as in eqs.(\ref{PdegenMagX},\ref{PdegenMagZ}). Here $ g_{\mu\nu}$ is the metric tensor and pressure $p(x^\mu)$ is a scalar. The covariant components of  gravitational stress-energy tensor, which replaces the Einstein tensor in linearized gravity, can be expressed with the gravitational potential $\Phi_g$ as: $(\underset{g}{T})\, _{\mu\nu}=\frac{1}{4\pi G}\left(\frac{\partial \Phi_g}{\partial x^\mu}\frac{\partial \Phi_g}{\partial x^\nu}-g_{\mu\nu}\frac{\partial \Phi_g}{\partial x^\lambda}g^{\lambda\sigma}\frac{\partial \Phi_g}{\partial x^\sigma}\right)$, and the contravariant  components of electromagnetic stress energy tensor can be expressed with electromagnetic vector potential 1-form  ($\cal A$) as: $(\underset{m}{T})^{\mu\nu}=\star\left[({\bf d}x^\mu\wedge{\bf d}{\cal A})\wedge\star({\bf d}x^\nu\wedge {\bf d}{\cal A})\right]-\frac{1}{2}g^{\mu \nu}({\bf d}{\cal A}\wedge \star {\bf d}{\cal A})$. The dynamic equilibrium condition can be expressed in Minkowski coordinates ($t,\,x\, ,y,\, z$) in vector form as: $(\underset{g}{T})^{\mu\nu}._\nu\rightarrow (\frac{1}{4\pi G}\Delta\Phi_g)\nabla \Phi_g=\rho \nabla \Phi_g$;   $\quad (\underset{m}{T})^{\mu\nu}._\nu\rightarrow -\rho_e\nabla \Phi_e-{\vec j}_e\times \vec B$, where $\rho_e$ and ${\vec j}_e$ are charge and current density and $\Phi_e$ is electric potential;  $(\underset{p}{T})^{\mu\nu}._\nu\rightarrow \nabla p$. 

To obtain analytic insight into  interaction of magnetic field with gravity,   we again return to the simplest polytropic model, the one for $n=0$. The basic equations  (\ref{RotConstr}) and  (\ref{LErot}) with $n=0$ and $\omega=0$ still apply together with the Poisson equation for gravitational field, however,  boundary conditions change. The incompressible fluid model makes boundary conditions particularly simple, because the magnetic surface may (almost) coincide with the surface of the star to make the atmosphere  infinitely thin. We  show that in this case an oblate elliptical surface satisfies the stress energy condition. 

In elliptic coordinates \ref{EliptCOO}, the internal and external gravitational potentials (\ref{FiGNot},\ref{FiGZun}) can be written as: 
\begin{align}
\Phi_g=&\frac{1}{3}a^2\Omega^2\left (\xi^2-1-2(1+e^2)\frac{\arctan e}{e}+\left(\frac{e-(1+e^2)\arctan e}{e}\right. \right.\nonumber \\&\left. \left.+\frac{e(3+2e^2)-3(1+e^2)\arctan e}{e^3}\xi^2\right)P_2(\cos \theta)\right)
 \qquad \xi\leq 1 \nonumber \\
\rightarrow & \frac{1}{3}a^2\Omega^2 \left(-3+\xi^2+e^2\left(-\frac{4}{3}+\frac{2}{15}(3\xi^2-5)P_2(\cos \theta)\right)\right)         +{\cal O}(e^4)\nonumber\\
\Phi_g=&-\frac{1}{3}a^2\Omega^2\left (2(1+e^2)\frac{\arctan\frac{e}{\xi}}{e}+\qquad\qquad\qquad\qquad\right.\nonumber\\ &\left.\qquad\qquad (1+e^2)\frac{(3\xi^2+e^2)\arctan\frac{e}{\xi}-3e\,\xi}{e^3}P_2(\cos \theta)\right)\quad\xi\geq 1 \label{GravMag}\\
\rightarrow & \frac{2}{3}a^2\Omega^2 \left(-\frac{1}{\xi}-\frac{e^2}{3\xi^3}\left( -1+3\xi^2+\frac{2}{5}P_2(\cos \theta)\right) \right)   +{\cal O}(e^4) 
 \nonumber
\end{align}
The electromagnetic field of an ellipsoidal dipole can best be expressed  in terms of the vector potential one form. The main ingredient of the external field must be the static  magnetic dipole field satisfying the field equation $\star{\bf d}\star{\bf d}{\cal A}=0$. However, the discussion of chapter 3 suggests that a (small) electric quadrupole component ($A_0\, {\bf d}t $) ,  generated by the surface charge, must also be present. The same discussion suggests that an internal electric polarization is required to allow electrons to occupy nonzero angular momentum orbits. Thus, the internal solution must combine a pure magnetic field, an electric component representing the chage imbalance, and a sourceless electric quadrupole  generating the surface charge which makes the system electrically neutral.  The field, which satisfies the above  propositions, can be expressed as follows:
\begin{align}
{\cal A}=&-\frac{B_0}{2}\,a^2(\xi^2 +e^2)\sin^2\theta\,{\bf d}\phi \nonumber\\&+S\,a^2(\xi^2-1)\left(1+\frac{2e^2}{3+e^2}P_2(\cos \theta)\right)\frac{c{\bf d}t}{a}  \nonumber \\&+Q a^2\left(\xi^2+\frac{e^2}{3}\right)P_2(\cos \theta)\frac{c{\bf d}t}{a}\qquad\quad\qquad\quad\qquad \qquad\qquad \qquad \xi\leq 1\nonumber\\
{\cal A}=&-\frac{B_0}{2}a^2(1+e^2)\frac{(\xi^2+e^2)\arctan\frac{e}{\xi}-e\,\xi}{(1+e^2)\arctan e-e}\sin^2\theta\,{\bf d}\phi \qquad\,\nonumber\\&+Q a^2(1+\frac{e^2}{3})\frac{(3\xi^2+e^2)\arctan\frac{e}{\xi}-3e\,\xi}{(3+e^2)\arctan e-3e}P_2(\cos \theta)\frac{c\,{\bf d}t}{a} \qquad\qquad\,\xi\geq 1
\label{MagMag}
\end{align}
 
In the next step we express stress energy tensors $(\underset{g}{T})_{\mu \nu}$, $(\underset{m}{T})_{\mu \nu}$ and $(\underset {p}{T})_{\mu \nu}$. The calculation of the first two is straightforward but quite messy and needs not be reproduced here in full detail, except for the fact that the divergence of the electromagnetic stress energy tensor can be written as:
\begin{equation}
(\underset{m}{T})_\mu^{\;\;\nu}._\nu=\frac{18 S^2}{\mu_0(3+2e^2)}\begin{pmatrix} 0\\-\xi(2+e^2+e^2\cos2\theta)\\e^2(1-\xi^2)\sin 2\theta\\0\end{pmatrix}-\frac{3 Q\,S}{\mu_0}\begin{pmatrix}0\\- \xi(1+3\cos 2\theta)\\(3\xi^2+e^2)\sin 2\theta 
\\0 \end{pmatrix}
 \label{EMpressContr}
\end{equation}
which is a gradient of a scalar  expressed in the formalism of differential forms as:
\begin{align}
{\bf d}x^\mu(\underset{m}{T})_\mu^{\;\;\nu}._\nu=\frac{1}{\mu_0}{\bf d}&\left[\frac{9S^2}{3+e^2}\left((2+e^2)\xi^2+e^2(\xi^2-1)\cos(2\theta)\right)\right. \nonumber\\&\left.+\frac{3 Q\,S}{2}\left(\xi^2+(3\xi^2+e^2)\cos(2\theta)\right)\right]\, .
\end{align}
The interior equilibrium condition can, therefore, be expressed only by the exterior derivative of a 0-form:
\begin{align}
&{\bf d}x^\mu(\underset{g}{T})_\mu^{\,\,\nu}._\nu+{\bf d}x^\mu(\underset{p}{T})_\mu^{\,\,\nu}._\nu+{\bf d}x^\mu(\underset{m}{T})_\mu^{\,\,\nu}._\nu=\nonumber \\&{\bf d}(\rho \Phi_g)+{\bf d}p+
\frac{1}{\mu_0}{\bf d}\left[\frac{9S^2}{3+e^2}\left((2+e^2)\xi^2+e^2(\xi^2-1)\cos(2\theta)\right)\right .\nonumber \\  &\qquad\qquad\qquad\qquad\left.+\frac{3 Q\,S}{2}\left(\xi^2+(3\xi^2+e^2)\cos(2\theta)\right)\right]\, ,
\label{PressDef}
\end{align}
which must vanish. This equation defines pressure up to a constant $p_1$. One notes, however, that the interior condition can still be satisfied if a divergence-free tensor is added to pressure. By adding  $\beta\,\frac{a^2\Omega^4}{\pi G}(\underset{zz}T)$, where  $(\underset{zz}T)_{\mu\nu}{\bf d}x^\mu{\bf d}x^\nu={\bf d}z^2$, one can model magnetically induced pressure anisotropy (cf. Eqs \ref{PdegGX} and \ref{PdegGZ}), where $\beta$ is a dimensionless constant determining its strength.    

 Boundary conditions (\ref{BdCondMR}) that remain to be solved are: $\Delta T^{\mu \xi}=\left(\underset{m}{T}^{\mu\xi}\right)_{ext}-\left(\underset{p}{T}^{\mu\xi}+\underset{m}{T}^{\mu\xi}\right)_{int}=0$ for $\mu\rightarrow \xi$ and $\mu\rightarrow \theta$, while the other two components are identically zero. We introduce dimensionless variables
\begin{align}
&p_1=\rho a^2\Omega^2\,\pi_1,&  B_0=a\, \Omega^2\sqrt{\frac{\mu_0}{\pi\,G}}\,{\cal B}\, ,\label{DimLess}\\ & Q=a\, \Omega^2\sqrt{\frac{\mu_0}{\pi\,G}}\,{\cal Q},& S=a\, \Omega^2\sqrt{\frac{\mu_0}{\pi\,G}}\,{\cal S}\, ,\nonumber
\end{align}
and obtain the following expression for the $(\xi,\theta)$ component:
\begin{equation}
\Delta T_{\xi\theta}=\left(c_0+c_2 \cos(2\theta)\right)\sin(2\theta)\, ,
\label{BCKsiTheta}
\end{equation}
where:
\begin{align}
c_0=&-\frac{6e^3(1+e^2)}{(1+e^2)\arctan e - e}{\cal B}^2-\frac{6e^5+2e^7}{(3+e^2)\arctan e-3e}{\cal Q}^2\nonumber \\&-18(1+e^2)(2+e^2){\cal Q}\,{\cal S}+6(1+e^2)\beta\, ,\nonumber \\
c_2=&-\frac{6e^5(3+e^2)}{(3+e^2)\arctan e-3e}{\cal Q}^2-18(e^2+e^4){\cal Q}{\cal S}\, .
\label{CKOF}
\end{align}
The $(\xi,\xi)$ component of the boundary condition is expressed in the form:
\begin{equation}
\Delta T_{\xi\xi}=\frac{1}{2}d_0+\frac{5}{2}d_2 P_2(\cos \theta)+\frac{9}{2}d_4  P_4(\cos \theta)\, ,
\label{BCThetaTheta}
\end{equation}
with the coefficients:
\begin{align}
d_0&=-\frac{8e^3(1+e^2)(2e+e^3-2(1+e^2)\arctan e)}{3((1+e^2)\arctan e -e)^2}{\cal B}^2\nonumber\\&+\frac{64 e^5(9e+9e^3+e^5-3(3+4e^2+e^4)\arctan e)}{45((3+e^2)\arctan e-3e)^2}{\cal Q}^2 +\nonumber \\
&\frac{8e^2(19+29e^2+13e^4+3e^6)}{5(3+e^2)}{\cal Q}{\cal S}\nonumber\\&+\frac{24(1+e^2)^2(60+85e^2+34e^4+5e^6)}{5(3+e^2)^2}{\cal S}^2+\nonumber \\ &\frac{4}{3}(1+e^2)(3+e^2)\pi_1+\frac{8}{3}(1+e^2)^2\beta\nonumber\\&+\frac{8(1+e^2)^2(2(3+e^2)\arctan e-e)}{15e}
\end{align}
\begin{align}
d_2&=\frac{8e^3(1+e^2)(2e+e^3-2(1+e^2)\arctan e)}{15((1+e^2)\arctan e-e)^2}{\cal B}^2\nonumber\\&+\frac{128e^5(9e+9e^3+e^5-3(3+4e^2+e^4)\arctan e)}{315((3+e^2)\arctan e-3e)^2}{\cal Q}^2\nonumber\\&
+\frac{16(1+e^2)(21+11e^2+14e^4+6e^6)}{35(3+e^2)}{\cal Q}{\cal S}\nonumber\\&+\frac{48e^2(1+e^2)^2(14+23e^2+7e^4)}{35(3+e^2)^2}{\cal S}^2\nonumber\\&+\frac{8e^2}{15}(1+e^2)\pi_1+\frac{16}{15}(1+e^2)^2\beta\nonumber \\&+\frac{4(1+e^2)^2((21+18e^2+13e^4)\arctan e-21e-11e^3)}{105 e^3}
\end{align}
\begin{align}
d_4&=\frac{128e^5(9e+9e^3+e^5-3(3+4e^2+e^4)\arctan e)}{315((3+e^2)\arctan e-3e)}{\cal Q}^2\nonumber \\&+\frac{64e^2(3+3e^2+e^4+e^6)}{105(3+e^2)}{\cal Q}{\cal S}-
\frac{128 e^4(1+e^2)^2}{35(3+e^2)^2}{\cal S}^2\nonumber \\ &+\frac{16(1+e^2)^2((3+e^2)\arctan e-3e)}{315 e}\, .
\label{DKOF}
\end{align}
The five coefficients ($c_0, c_2, d_0, d_2$ and $d_4$) must vanish, which gives five nonlinear equations for the five constants  $\cal B$, $\cal Q$, $\cal S$, $\pi_1$ and $\beta$ . These complicated equations have surprisingly simple solutions.

 For $e=0$ the solution becomes: $\pi_1=-\frac{2}{3}(1+12{\cal S}^2), \, {\cal B}=0,\,{\cal Q}=0,\, \beta=0$. For ${\cal S}=0$ this is the clasical spherical gravitational solution. For ${\cal S}\neq 0$ the solution also contains an internal radial electric field (${\cal A}={\cal S}\,\Omega^2 \sqrt{\frac{\mu_0}{\pi\, G}}a^3(\xi^2-1)\frac{c{\bf d}t}{a}$) which is screened by negative surface charge, as discussed in the previous chapter. Noting that $A_0$ is the negative electrostatic potential, one can express the electric charge imbalance density as $-\varepsilon_0\,\Delta(A_0)=12\pi\sqrt{\varepsilon_0 G}\rho\, {\cal S}$.  Furthermore, by expressing mass density as $\rho= A_m\,m_p n_A$ and electron number density as $Z\,n_A$,  the ratio of charge imbalance density to electron number density can be written as $12\frac{A_m}{Z}(\sqrt{\pi\,G\,\varepsilon_0}\,\frac{m_p}{e})\,{\cal S}\approx 5.4\times 10^{-18}\frac{A_m}{Z}\,{\cal S}$.  

For nonzero ellipticity two types of solutions exist: ellipticity may be induced sollely by internal electric polarization:
\begin{align}
&\pi_1=-\frac{(1+e^2)(12+17e^2+10e^4+e^6)\arctan e-e(12+13e^3+3e^4)}{4e^5}\nonumber \\ &\quad\rightarrow\quad-\frac{14}{15}-\frac{27}{35}e^2+\frac{2}{105}e^4+\dots 
\end{align}

\begin{align}
{\cal S}&=\frac{(3+e^3)\sqrt{(3+e^2)\arctan e-3e}}{6e^2\sqrt{2 e}}\nonumber\\&\quad\rightarrow\quad\frac{1}{\sqrt{30}}-\frac{1}{21}\sqrt{\frac{2}{15}}e^2+\frac{1}{49}\sqrt{6}{5}e^4+\dots 
\end{align}
\begin{align}
{\cal B}&=0,\qquad\qquad {\cal Q}=0,\qquad\qquad \beta=0\, ,
\label{ElSol}
\end{align} 
or by a combination of electric polarization and magnetic field:
\begin{align}
\pi_1&=\frac{(6-e^2-7e^4)\arctan e-e(6+13e^2+9e^4)}{18e^3}\label{Pi1MagSol} \\& \quad\rightarrow\quad-\frac{8}{9}-\frac{217}{270}e^2+\frac{67}{945}e^4+\dots \nonumber
\\
\beta&=\frac{e(9+7e^2)-(9+10e^2+e^4)\arctan e}{18e^3}\nonumber \rightarrow \frac{4}{135}e^2-\frac{4}{189}e^4+\dots \\
{\cal B}&=\frac{1}{3}\frac{(1+e^2)\arctan e-e}{e^2}\rightarrow\quad \frac{2}{9}e-\frac{2}{45}e^3+\dots\label{BboundC}\\
{\cal S}&=\frac{1}{6}\\
{\cal Q}&=\frac{(1+e^2)((3+e^2)\arctan e)-3e}{2e^3(3+e^2)}\rightarrow -\frac{2}{45}e^2+\frac{8}{945}e^4+\dots
\label{ElMagSol}
\end{align}


The relation between ellipticity and magnetic field can be obtained also by applying the variational principle. We repeat the argument used in deriving  eq.\ref{RotHamPrinc} and replace rotational energy by magnetic energy as derived in section 3.1. The gravitational energy,  magnetic energy and the action can be written as:
\begin{align}
&W_g\quad=-\frac{3G \ M^2}{5 R_{ef}} \frac{(1+e^2 )^{1/3} \arctan e}{e}\nonumber\\
&W_{mag}=  \frac{B_0^2}{2 \mu_0} \left( \frac{4 \pi}{3} R_{ef}^3 \right) \frac{ (e-\arctan e )}{(1+e^2) \arctan e - e} \nonumber\\
&A=W_g+W_{mag}+\lambda \frac{1}{\mu_0}B_0 \left(\frac{4\pi}{3}R_{ef}^3\right)\, ,
\label{ActnMag}
\end{align}
where $R_{ef}=(1+e^2)^{1/6}a$.
Solving $\frac{\partial A}{\partial B_0}=\frac{\partial A}{\partial e}=0$ and expressing $B_0$ with dimensionless $\cal B$, one obtains \footnote{Note that during the variational process the volume and density of the star must remain constant, therefore, during this process,\hfill\break we introduce a new dimensionless magnetic field with $B_0=R_{ef}\Omega^2\sqrt{\frac{\mu_0}{\pi\,G}}\tilde{\cal B}$.}:
\begin{equation}
{\cal B}=\sqrt{\frac{2}{15}}\frac{(1+e^2)\arctan e-e}{e^2}
\label{Bact}
\end{equation}
Comparing eqs. \ref{Bact} and \ref{BboundC} one finds the same functional dependence of $\cal B$ on $e$, only the factor $\sqrt{\frac{2}{15}}$ above is about 10\% higher than $\frac{1}{3}$ from eq. \ref{BboundC}. 
Noting  that in the polytropic model  the scale of ${\cal B}$ is inversely proportional to central pressure: ${\cal B}=\sqrt{\frac{\pi G}{\mu_0}}\frac{1}{a\,\Omega^2}=\frac{1}{\sqrt{\mu_0\pi\,G}p_c}B_0$, where $p_c$ is the central pressure as defined by polytropic model (cf derivation of eq.\ref{LEn0}) the discrepancy may be plausible, since in the variational problem we did not take into account that the equation of state (\ref{PressDef}) is modified by additional pressure generated by electric polarization in the presence of electromagnetic field. 
 
\subsection{Magnetic surface instability}
\label{MGSI}
 In the previous section we have shown that a star can reach equilibrium with magnetic field if it assumes elliptic shape. Yet, the above analysis could not prove that the open boundary gravito-magnetic problem has a unique solution. Experiments with magnetic fluid in the presence of magnetic field suggest that the surface of magnetic fluid develops a spiny structure which disperses a simple dipolar field into a high order mulipole field which shields the internal field from spreading away from the source. A detailed analysis of this phenomenon presents a rather difficult problem. For the present analysis we assume, as in the $n=0$ model, that the magnetic surface coincides with the physical surface of the star. To get some idea how the magnetic surface can be deformed by forming wrinkles, we ask  what other  surface  shapes can also minimize the energy of the star with a given magnetic moment.   
 
In principle any infinitesimally deformed sphere  can be described in spherical coordinates by expanding the radius in terms of spherical harmonics as $r_S(\theta,\phi)=R+\sum_{l,m}a_{l,m}Y_l^m(\theta,\phi) $.  The deformation generally increases gravitational energy, while magnetic energy for a given magnetic moment decreases, if the  magnetic field of a dipole is dispersed into higher multipoles. Higher multipoles confine the magnetic field more tightly to the vicinity of the surface, reducing the volume of space filled with magnetic flux and therefore lowering the external magnetic energy for a fixed magnetic moment. If magnetic pressure can do enough work during the deformation to compensate for the increase of gravitational energy, then such a deformation leads to a lower energy state. 

Because of complexity of the general problem, we limit the analysis to cylindrically symmetric deformations of the form $r_S(\theta)=a\left(1+\delta_\lambda P_\lambda(\cos \theta)\right)$ for even positive integers $\lambda$.  
We solve gravitational and electromagnetic equations in a system of coordinates $\xi,\, \zeta,\,\phi$  defined so that $\zeta=\cos \theta$ and  $x=r_S(\zeta)\,\xi\,\sqrt{1-\zeta^2}\,\cos \phi,\,y=r_S(\zeta)\,\xi\,\sqrt{1-\zeta^2}\,\sin \phi,\,z=r_S(\zeta)\,\xi \,\zeta$, such that $\xi=1$ represents the surface. 
Functions $\psi^{(-)}_l=\xi^l r_S(\zeta)^lP_l(\zeta)$ and $\psi^{(+)}_l=\frac{1}{\xi^l r_S(\zeta)^l+1}P_l(\zeta)$ solve the Poisson equation for $\xi<1$ and $\xi>1$ respectively, while 1-forms $\alpha^{(-)}_l=(1-\zeta^2)\xi^{l+1}r_S^{l+1}(\zeta)\frac{dP_l(\zeta)}{d\zeta}{\bf d}\phi$ and $\alpha^{(+)}_l=\frac {1-\zeta^2}{\xi^lr_S^l}\frac{dP_l(\zeta)}{d\zeta}{\bf d}\phi$ solve the electromagnetic equation ${\bf d}\star{\bf d}\alpha=0$. Together with the particular solution of gravitational Poisson equation and these functions one can construct solutions of gravitational and magnetic field equivalent to eqs.(\ref{GravMag}) and (\ref{MagMag}). The only difficulty is that  eigenfunctions $\psi^{(-)}_l,\,\psi^{(+)}_l$ and 1-forms $\alpha^{(-)}_l,\,\alpha^{(+)}_l$ do not have the same angular dependence at the surface, since $r_S$ is a function of $\zeta$. Therefore, matching interior to exterior solution requires expanding  eigenfunctions at the surface  in a common basis of Legendre polynomials. We use the first order expansion in terms of parameter $\delta$. Numerical procedure is straightforward, but quite tedious and must be done numerically. 

Numerical  results lead to expansion coefficients $k^{(g)}_\lambda$ and $k^{(m)}_\lambda$ through which  energy is expressed as a function of parameter $\delta$  in the form
\begin{align}
W_g&=-\frac{3}{5}\frac{G\,M^2}{R_{ef}}(1-k^{(g)}_\lambda\,\delta^2)\label{GravEnDef}\\
W_m&=\frac{3}{4}\frac{G\,M^2}{R_{ef}}{\cal B}^2\left(1+k^{(m)}_\lambda\,\delta^2\right)\, .
\label{MagEnDef}
\end{align} 
\begin{figure}[h]
\includegraphics[width=8.5cm]{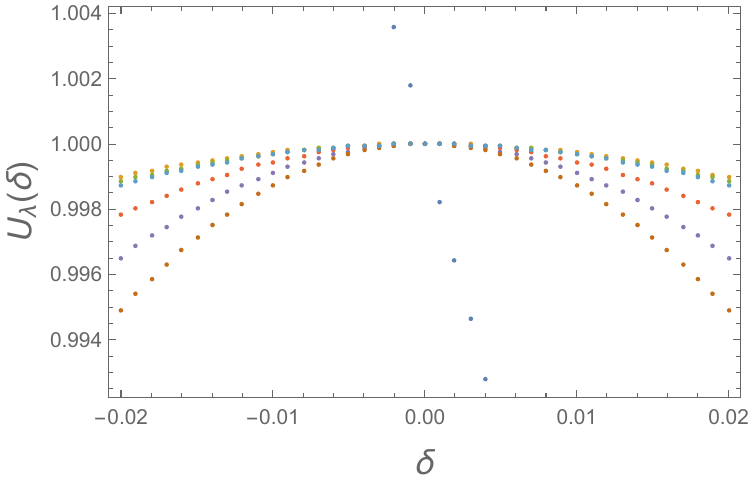}
\caption{Numerical results of calculating magnetic energy $W_{m, \lambda}(\delta)$ to determine coefficients $k_\lambda^{(m)}(\delta)$ for $\lambda=4,\,8,\,10,\,12,\,14,\,30$. The steep linear slope corresponds to 
 $U_2^{(m)}(\delta)\propto \left(1-\frac{9}{5}\,\delta\right)$.}
\label{MPoliE}
\end{figure} 

The case $\lambda=2$ is particular. Here the sphere is deformed into a prolate ellipsoid for small positive $\delta$ and an oblate one for negative $\delta$. This  solution naturally leads to results of  previous section noting that magnetic moment varies with $\delta$ and that $\delta=(2 (-1 + \sqrt{1 + e^2}))/(2 + \sqrt{1 + e^2})\sim \frac{e^2}{3}-\frac{5e^4}{36}+\dots$. 
\begin{figure} 
\centering
\includegraphics[width=.45\textwidth]
{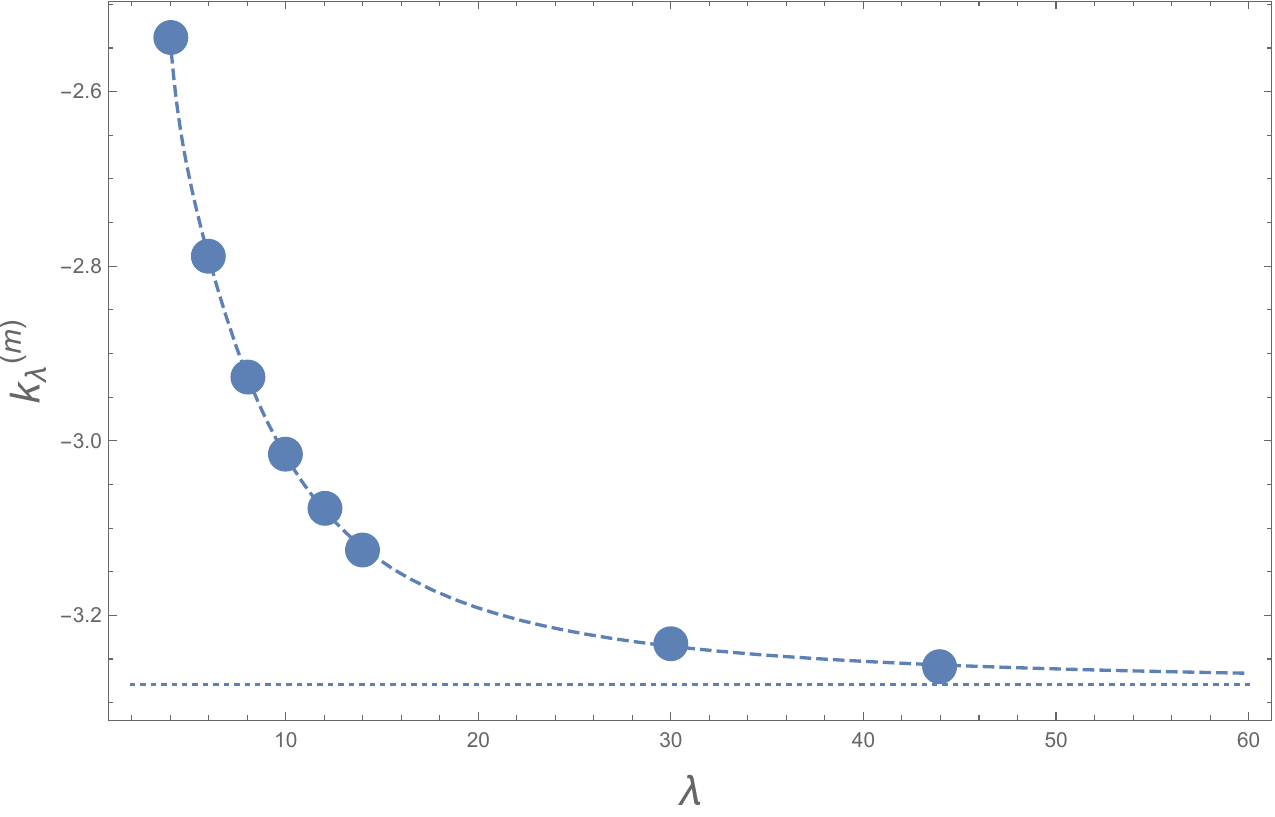}\hfill
\includegraphics[width=.45\textwidth]
{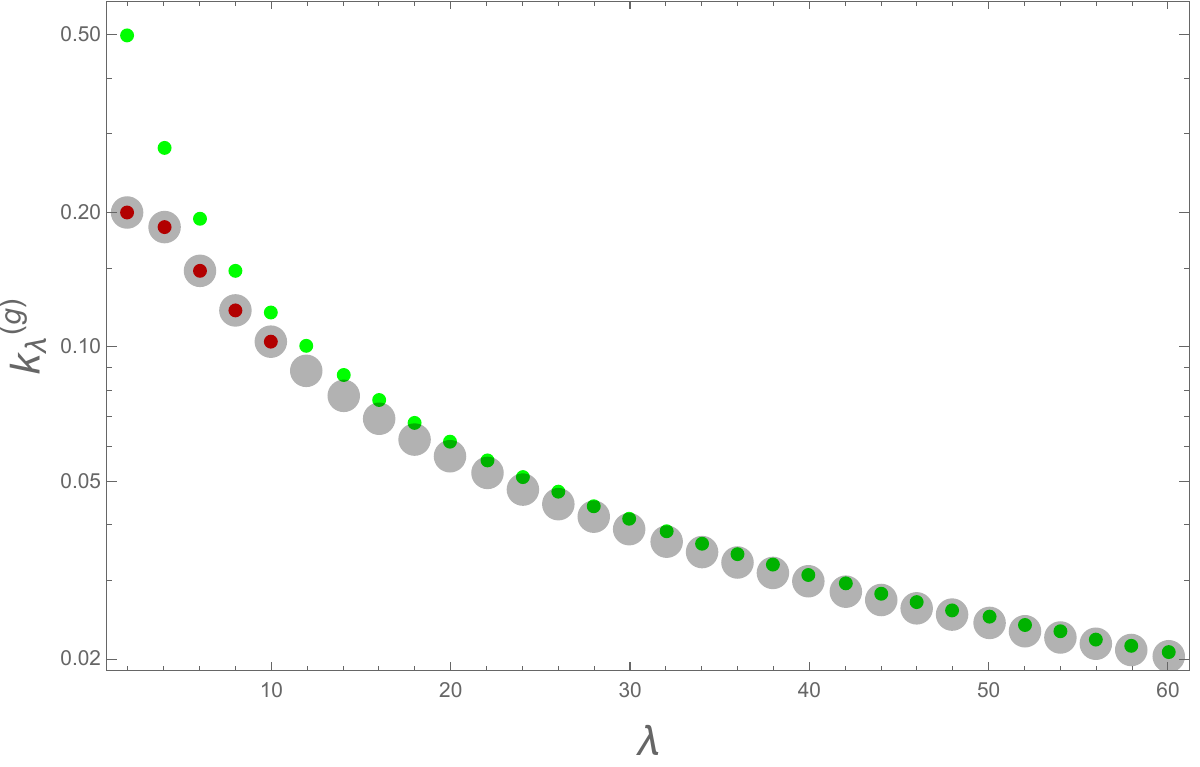}
\caption{left: numerical values of  $k_\lambda^{(m)}$ (blue points) and fit with inverse powers of $\lambda$;\hfill\break right: $k_\lambda^{(g)}$  numerical values (magenta), asymptotic approximation (green), $k_\lambda^{(g)}=\frac{5}{2(2\lambda+1)}\left(1-\frac{3}{2\lambda+1}\right)$ (grey)}
\label{Koefkmg}
\end{figure}

For all $\lambda>2$ both energies have an extremum at $\delta=0$ - a minimum for gravitation and a maximum for magnetism. If $\left(\frac{3}{5}k_\lambda^{(g)}+\frac{3}{4}k_\lambda^{(m)}{\cal B}^2\right)\delta^2>0$ then gravity prevails, and the energy has a minimum which makes the system  stable against perturbation, otherwise the chosen deformation allows a lower energy state making the system  unstable with  respect to this perturbation.

\newpage

The numerically calculated coefficients $k_\lambda^{(m)}$ and $k_\lambda^{(g)}$ are shown in Fig.\ref{Koefkmg} together with extrapolations for large\footnote{The expression $k_\lambda^{(g)}=\frac{5}{2(2\lambda+1)}\left(1-\frac{3}{2\lambda+1}\right)$ fits the first five numerically obtained values to nine significant figures and approaches asymptotically the $1/\lambda$ dependence which follows from first order perturbation analysis. In the case of magnetic coefficient $k_\lambda^{(m)}$ we don't have a firm argument for asymptotic behaviour, except for the fact that calculated parameters can be quite accurately fit with a Taylor series of inverse powers of $\lambda$.}  $\lambda$. Since $k_\lambda^{(m)}$ approaches a constant value ($\sim3.27$) for large $\lambda$ and $k_\lambda^{(g)}$ decreases in absolute value,  the magnetic surface is unstable with respect to  deformations with $\lambda $ greater than the solution of equation ${\cal B}=\frac{2\sqrt{\lambda-1}}{\sqrt{k_\lambda^{(m)}}(2\lambda+1)}\sim\frac{1}{\sqrt{3.27 \lambda}}$. 

Stellar magnetic fields are typically quite small, which means that magnetic surface can only be unstable with respect to deformations of high order
\begin{equation}
\lambda_{crit}\approx\frac{1}{\vert k_{\lambda\rightarrow\infty}^{(m)}\vert\,{\cal B}^2 }\, ,
\label{LamKrit}
\end{equation}
so that deformations corrugate the surface on a small length scale. 

Predicting the  shape  of such corrugations is a difficult task, which is beyond the scope of this article. However, a similar Rosensweig instability \citep{1987AnRFM..19..437R} is observed in experiments with ferrofluids whose surface in a jar becomes vertically corrugated if a sufficiently strong magnet is approached from below. 

Let us  calculate the typical size of corrugations that might be expected on a magnetic object the size of Earth. Expressing $\cal B$ with polar field $B_0$ according to eq.(\ref{DimLess}), we obtain $\lambda_{crit}\approx\frac{\mu_0a^2\Omega^4}{\pi\,G\,k_\lambda^{(m)}B_0^2}\sim\frac{4\times 10^5}{\left(B_0/\mathrm{Tesla}\right)^2}$, which is to say that the magnetic field of $\sim 100\,$ gauss would generate corrugations of about $2\pi R_{Earth}/\lambda_{crit}\sim$ 1$\mathrm{cm}$ . This is roughly comparable to the size of ripples observed in experiments with magnetic fluid. 
Corrugation of magnetic surface may appear unimportant regarding the magneto-mechanical equilibrium of a star, which is dominated by the magnetic dipole moment defined by the integral of magnetization over the volume of the star. However, stellar magnetic fields are usually deduced from the dipole moment deduced from the external magnetic field. These two dipole moments are the same only for spherical magnets. Corrugations disperse magnetic energy  between high multipoles, which makes the exterior dipole component weaker than the one defined by interior magnetization.   

\begin{figure}[h]
\includegraphics[width=8.5cm]{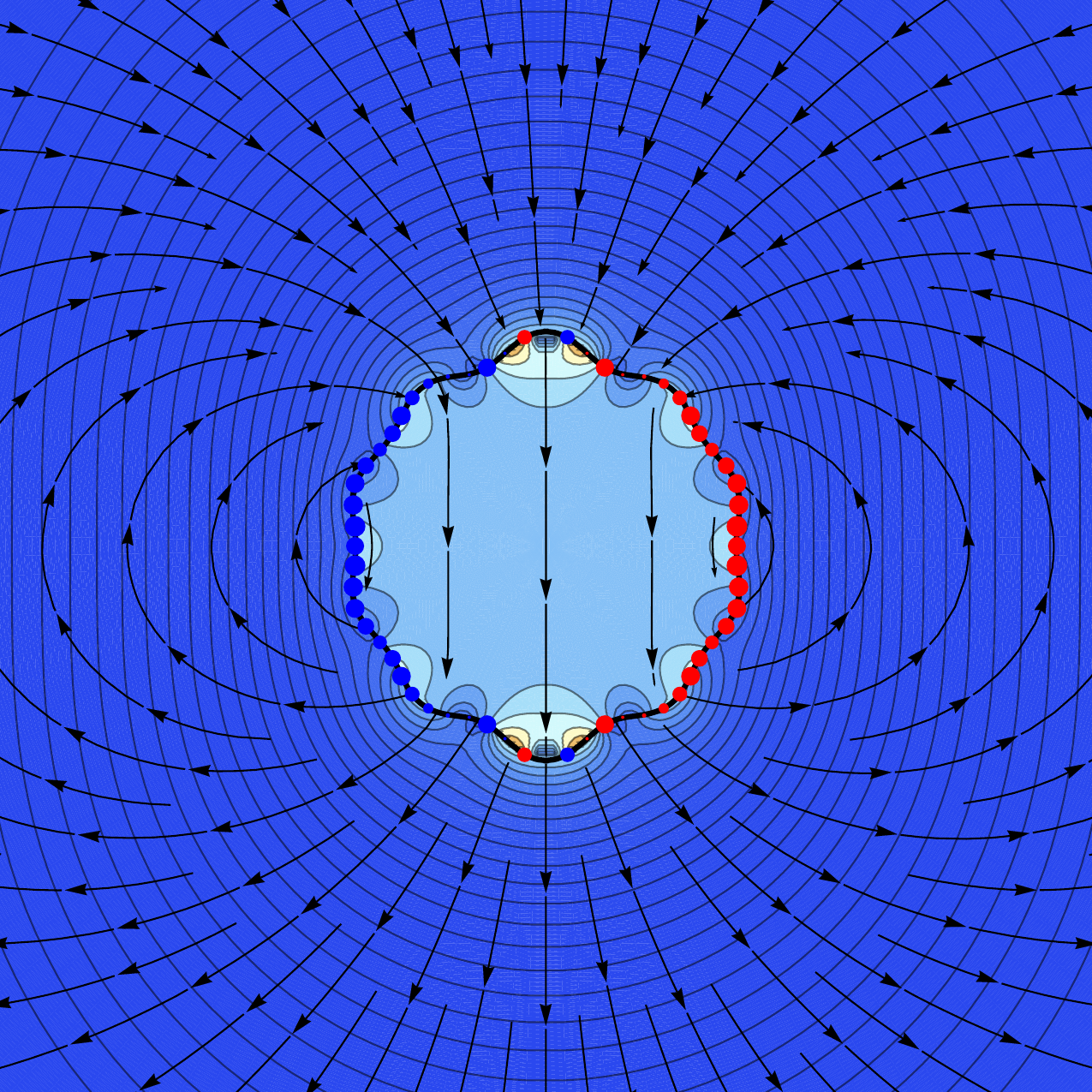}
\caption{Magnetic field of a spherical magnet deformed with $r_S=1+0.1\,P_{10}(\cos \theta)$. Magnetic energy density contours and red and blue dots representing the surface current are coded as in Fig.\ref{SL1a}}
\label{10FoldBuckling}
\end{figure}

The phenomenon is illustrated in Fig.\ref{10FoldBuckling}, which represents magnetic field of a solid ideal ferromagnet whose surface is described by $r_S=1+0.1\,P_{10}(\cos \theta)$. Corrugations generate additional mini-poles which redistribute the internal field to the 10-pole and other higher multipoles. \newpage The details of a free surface of a star are much more complicated, because we do not know the shape and the length of corrugations. The fact that the magnetic surface must be in most cases below the actual surface of the star complicates the matter even further. Yet it seems quite safe to say that the internal magnetic field of stars is stronger, if not much stronger than the field deduced from  measurements, which are usually based on the assumption of magnetic dipole. 

Finally, we must add a caveat. The above analysis is valid only if the magnetic surface is at the physical surface of the star. More realistic models must take into account the fact that, a given magnetization can be sustained inside the star, only up to a certain level to which the state of matter is adequately represented by equations (\ref{DegenEMag},\ref{PdegenMagX}, \ref{PdegenMagZ}) in the domain of pressure domination (cf. Fig.(\ref{MagEnDensP})). Magnetic surface must be at the level where electron density falls below the critical density at which the degenerate gas changes from pressure to magnetic field dominated. Beyond this level pressure is magnetic field dominated, but density is not. Therefore, magnetic pressure does not push against gravity. As a result the magnetic surface can deform without much change in gravitational energy. Therefore, one should expect that the magnetic instability sets in at much, much larger wavelengths than predicted by eq.\ref{LamKrit}. 


\section{Rotating magnetic star}
\label{RotMagSt}

The extension of "Lane-Emde equation" for a magnetic star to a rotating magnetic star is straightforward. Rotation adds "nondiagonal" terms to the matter tensor and  centrifugal pressure ($\frac{1}{2}\rho\,\omega^2R^2$) to pressure. In Minkowski coordinates the stress tensor of moving matter is written as: $\underset{p}{T}^{\mu\nu}=\frac{\rho c^2+p}{c^2}u^\mu u^\nu+p \eta^{\mu \nu}$, where $u^\mu$ are components of 4-velocity and $\eta^{\mu\nu}$ represents the Minkowsky tensor. This form is the standard stress tensor of a rigidly rotating fluid in the linerized (Newtonian) limit, the velocity components of matter rotating about the $z$ axis are: $v_x=-\omega\,y$, $v_y=\omega\,x$, so that the components of rotating matter stress-energy tensor are: 
\begin{align}
\left(\underset{p}{T}\right)_{\mu\nu}&\longrightarrow 
\begin{pmatrix}-\rho\,c^2+p&0&0&0\\0&p&0&0\\
0&0&p&0\\
0&0&0&p
\end{pmatrix}\nonumber 
 + \begin{pmatrix} \frac{1}{2}\rho\,\omega^2(x^2+y^2)&
-\rho c\,\omega\,y
&\rho c\,\omega\,x &0\\-\rho c\,\omega\,y&\rho\,\omega^2\,y^2&-\rho\,\omega^2\,x\,y & 0 \\ \rho c\,\omega\,x&
-\rho\,\omega^2\,x\,y&\rho\,\omega^2\, x^2&0\\
0&0&0&0 
\end{pmatrix}
\nonumber
\end{align}
The divergence of the second tensor in the above expression  $\underset{rot}{T}$ only has space-like components and can be written as $\left(\underset{rot}{T}\right)^{\mu\nu}._\nu\longrightarrow \rho\, \omega^2\vec R+\rho\, c ( \frac{d{\vec\omega}}{dt} \times \vec R)$, where $\vec R$ stands for $\lbrace x,\,y,\,0\rbrace$. Here the centrifugal term contributes only to the diagonal spatial components, reflecting the additional pressure generated by rigid rotation. If $\frac{d\omega}{dt}=0$, the problem simplifies, since the space-time components of $\underset{rot}{T}$ contribute no force to rotating mater.  Such a simplification occurs only if the electromagnetic stress tensor has no space-time components to be compensated by matter, i.e.  if the magnetic field is aligned with rotation axis.  We restrict our analysis to the case where the magnetic field is aligned with the rotation axis, since any misalignment produces time-dependent stresses that can not be treated witin the stationary variational framework used here.  Our discussion of rotating magnetic stars is limited to this simple example.  

Expressing spatial components of $\underset{rot}{T}$  as a bilinear 1-form and translating it in the language of elliptic coordinates:
\begin{equation}
{\bf d}x^i\left(\underset{rot}{T}\right)_{ij}{\bf d}x^j=\rho\,\omega^2\,(x\, {\bf d}y-y\,{\bf d}x)^2=\rho\,\omega^2(\xi^2+e^2)^2\sin^4\theta\,{\bf d}\phi^2\, ,\nonumber
\end{equation}
one finds that the only nonzero component is $\underset{rot}{T^{\phi\phi}}=\rho\,\omega^2$.  
The equilibrium of the star is again expressed by eq.(\ref{inequil}),where gravitational and magnetic field tensors follow from (\ref{GravMag}) and (\ref{MagMag}), while  $\underset{p}{T}$ has the additional $\omega^2$ term. The equation \ref{PressDef} now turns into
\begin{align}
&{\bf d}x^\mu(\underset{g}{T})_\mu^{\,\,\nu}._\nu+{\bf d}x^\mu(\underset{p}{T})_\mu^{\,\,\nu}._\nu+{\bf d}x^\mu(\underset{m}{T})_\mu^{\,\,\nu}._\nu=
\label{PressDefR} \\&{\bf d}(\rho \Phi_g)+{\bf d}\left(p+\frac{\rho\,\omega^2}{2}(\xi^2+e^2)\sin^2\theta\right)\nonumber \\&+
\frac{1}{\mu_0}{\bf d}\left[\frac{9S^2}{3+e^2}\left((2+e^2)\xi^2+e^2(\xi^2-1)\cos(2\theta)\right)\right.\nonumber \\ &\left.\qquad\qquad\qquad +\frac{3 Q\,S}{2}\left(\xi^2+(3\xi^2+e^2)\cos(2\theta)\right)\right]
\nonumber
\end{align}
The boundary condition (\ref{BCKsiTheta}) is unchanged by rotation, so that the corresponding conditions (\ref{CKOF}) remain unchanged, while the conditions  (\ref{BCThetaTheta}) are changed to: $d_0\rightarrow d_0+\frac{4}{3}(1+e^2)^2(1+\frac{1}{5}e^2)\left(\frac{\omega}{\Omega}\right)^2$, $d_2\rightarrow d_2-\frac{4}{15}(1+e^2)^2(1-\frac{1}{7}e^2)\left(\frac{\omega}{\Omega}\right)^2$ and\hfill\break $d_4\rightarrow d_4-\frac{16}{315}e^2(1+e^2)^2\left(\frac{\omega}{\Omega}\right)^2$.  The solution of the new equations, equivalent to \ref{Pi1MagSol} - \ref{ElMagSol}, becomes:
\begin{align}
\pi_1&=\frac{(6-e^2-7e^4)\arctan e-e(6+13e^2+9e^4)}{18e^3}\nonumber \\ &-\frac{(24+35e^2+11e^4)\arctan e -e(24+31e^2+9e^4)}{18\left((3+e^2)\arctan e-3e\right)}{\tilde \omega}^2
\label{Pi1RotMagSol}
\end{align}
\begin{align}
\beta&=\frac{e(9+7e^2)-(9+10e^2+e^4)\arctan e}{18e^3}\nonumber\\&\quad+\frac{(9+10e^2+e^4)\arctan e-e(9+7e^2)}{18\left((3+e^2)\arctan e-3e\right)} 
\end{align}
\begin{align}
{\cal B}^2&=\left(\frac{1}{3}\frac{(1+e^2)\arctan e-e}{e^2}\right)^2-\frac{\left((1+e^2)\arctan e-e\right)^2}{9e\left((3+e^2)\arctan e-3e\right)}{\tilde \omega}^2\label{BboundCRot}
\end{align}
\begin{align}
{\cal S}^2&=\left(\frac{1}{6^2}-\frac{e^3}{36\left((3+e^2)\arctan e-3e\right)}{\tilde \omega}^2\right)
\end{align}
\begin{align}
{\cal Q}^2&=\left(\frac{1+e^2}{3+e^2}\frac{(3+e^2)\arctan e-3e}{2e^3}\right)^2\nonumber\\&-\left(\frac{1+e^2}{3+e^2}\right)^2\frac{(3+e^2)\arctan e-3e}{4e^3}{\tilde\omega}^2
\label{ElRozMagSol}
\end{align}
For the purely magnetic case (${\tilde\omega}=\frac{\omega}{\Omega}=0$) this solution goes to the solution \ref{Pi1MagSol} - \ref{ElMagSol}, while for pure rotation (${\cal B}=0$) it leads to the McLaurin solution \ref{omegaepsilon} and sets $\beta=0$, ${\cal S}=0$ and ${\cal Q}=0$.
\ref{ActnMag} as $W=W_g+W_{mag}+W_{rot}$, and $W_0$ is the gravitational energy of spherical star.

 Figure 9 shows a phase diagram, based on the exact $n$ = 0 polytropic solution, for stars with available measurements of rotation and magnetic field. The axes are the magnetic parameter $ {\cal B}$, the dimensionless spin parameter, $\tilde{\omega}$ and energy  excess $W$. We define $ {\cal B}$ as the ratio of the magnetic flux through the equatorial cross-section to the stellar mass, $\tilde{\omega}$ as the ratio of the observed stellar angular velocity, $\omega$, to the critical breakup value, $\Omega_{\mathrm{K}} = \sqrt{2 \pi G \rho} $, 
 and $W=W_g+W_{mag}+W_{rot}$ (cf. eq.\ref{EMWork}). A broad range of stellar objects is found to cluster in the same region of this diagram. This coincidence supports the interpretation that the electromagnetic field preserves its long-range character through magnetism.

 The Sun, Earth Jupiter and Saturn are represented  by blue points on dotted lines whose positions correspond to the measured angular velocity $\tilde \omega$. Parallel to those lines are lines (red for Sun, green for Earth, magenta for Jupiter and brown for Saturn) corresponding to their measured ellipticity parameter $e$. Note that calculated ellipticity parameters  for Earth and Jupiter are quite close to  measured values represented by coloured lines and can easily be understood within the polytropic model with $0<n<1.5$. However, the ellipticity of the Sun is less than even the softest polytropic model with  $n=3$ would predict and may even be correlated with solar cycle activity \citep{2019ApJ...875L..26I}. The rest of the diagram is mostly populated by pulsars, well known for their rotation and magnetic field. 
   \begin{figure*}[h!]
        \centering
        \includegraphics[width=15.5cm]{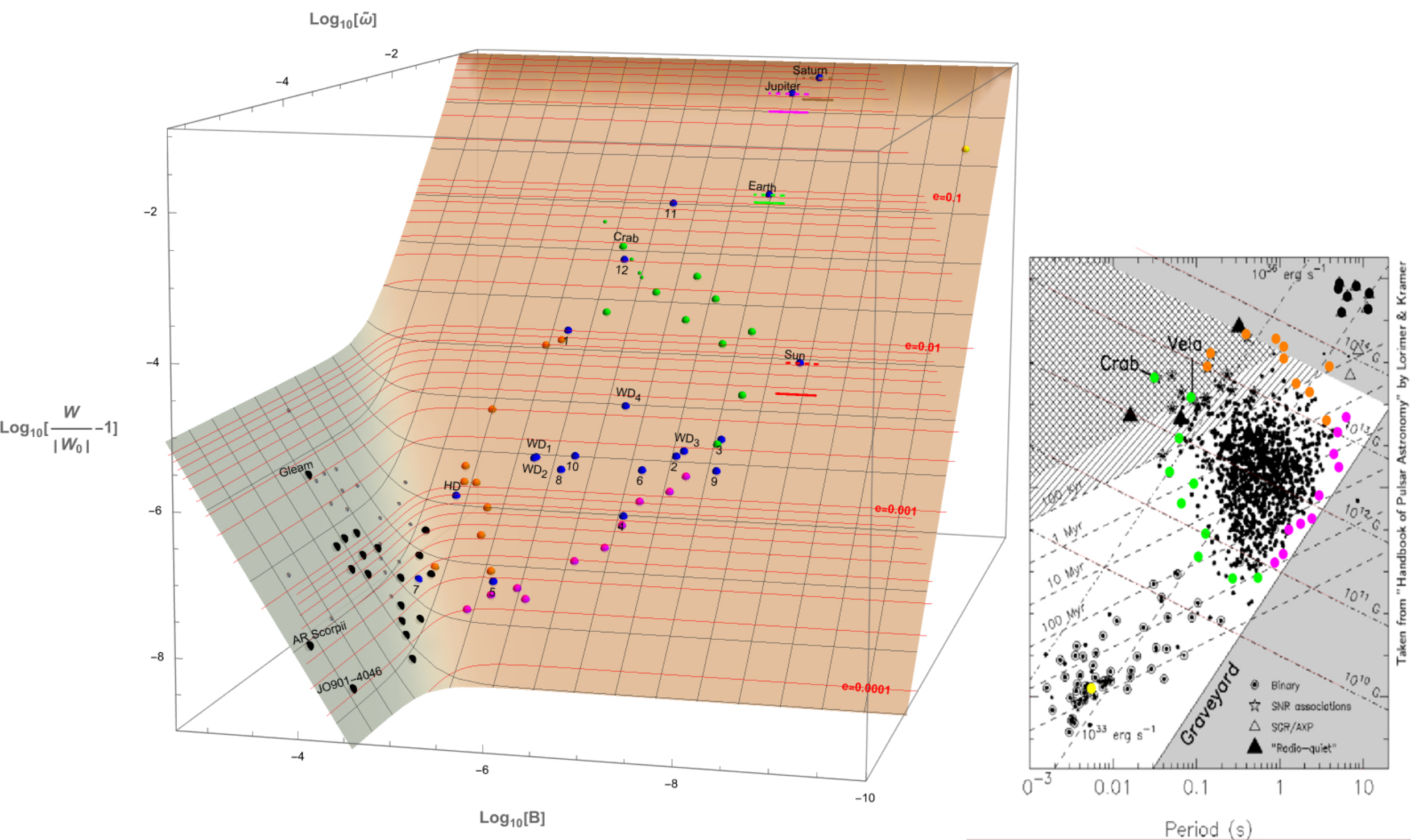}
            {\caption{Magnetism and rotation of the Sun, planets, pulsars, magnetars and white dwarfs. Colored pulsars from the border of the P-Pdot diagram on the right are represented with the same color in the left phase diagram (orange for high magnetic field pulsars, magenta for death line pulsars, green for young pulsars, and yellow for millisecond pulsars). The smaller satellites of the Crab pulsar represent its position on the diagram assuming its different masses from .67 to 2.09$M_\odot$ . Magnetars (shown in grey) are the only objects in this diagram with their elipticity controled by magnetic field. White dwarfs are represented as numbered ($\mathrm {WD}_1$ to $\mathrm {WD}_4$ and 1 to 12)  blue balls} \label{SLRotMagEn}}
    \end{figure*}

The positions of pulsars in the diagram were calculated from their positions in the P-Pdot diagram assuming their mass and radius to be 1.3$M_\odot$ and 13.5km respectively. To indicate the possible uncertainty in calculating $\tilde \omega$ and $\cal B$, the Crab pulsar has four small satellites, which were calculated by assuming pulsar masses to be 2.09, 2.03, 1.7, .67$M_\odot$, with respective  radii  (10, 12, 13, 14 kilometers), which  follow from a recent  neutron star mass-radius model \citep{2020Univ....6..119B}. The  pulsar population is represented by members of the boundary in the P-Pdot diagram shown at right of Fig. \ref{SLRotMagEn}; pulsars have the same color in both figures. Magnetars presented in  (\cite{2022NatAs...6..828C}) are also shown in black, together with their recently 
discovered unusual pulsar PSR J0901-4046  assuming that their mass is the largest possible neutron star mass 2.13$M_\odot$ according to \citep{2020Univ....6..119B}. The same magnetars are shown as smaller grey balls if their mass is assumed to be 1.3$M_\odot$.
According to \cite{2023PhRvD.107h1301S}  the magnetic field of  PSR J0901-4046 may be more than 100 times stronger, which would bring it to the level of "Gleam" (GLEAM-X J162759.5-523504.3  \cite{2022NatAs...6..828C}).

   \begin{figure*}[h!]
        \centering
        \includegraphics[width=12cm]{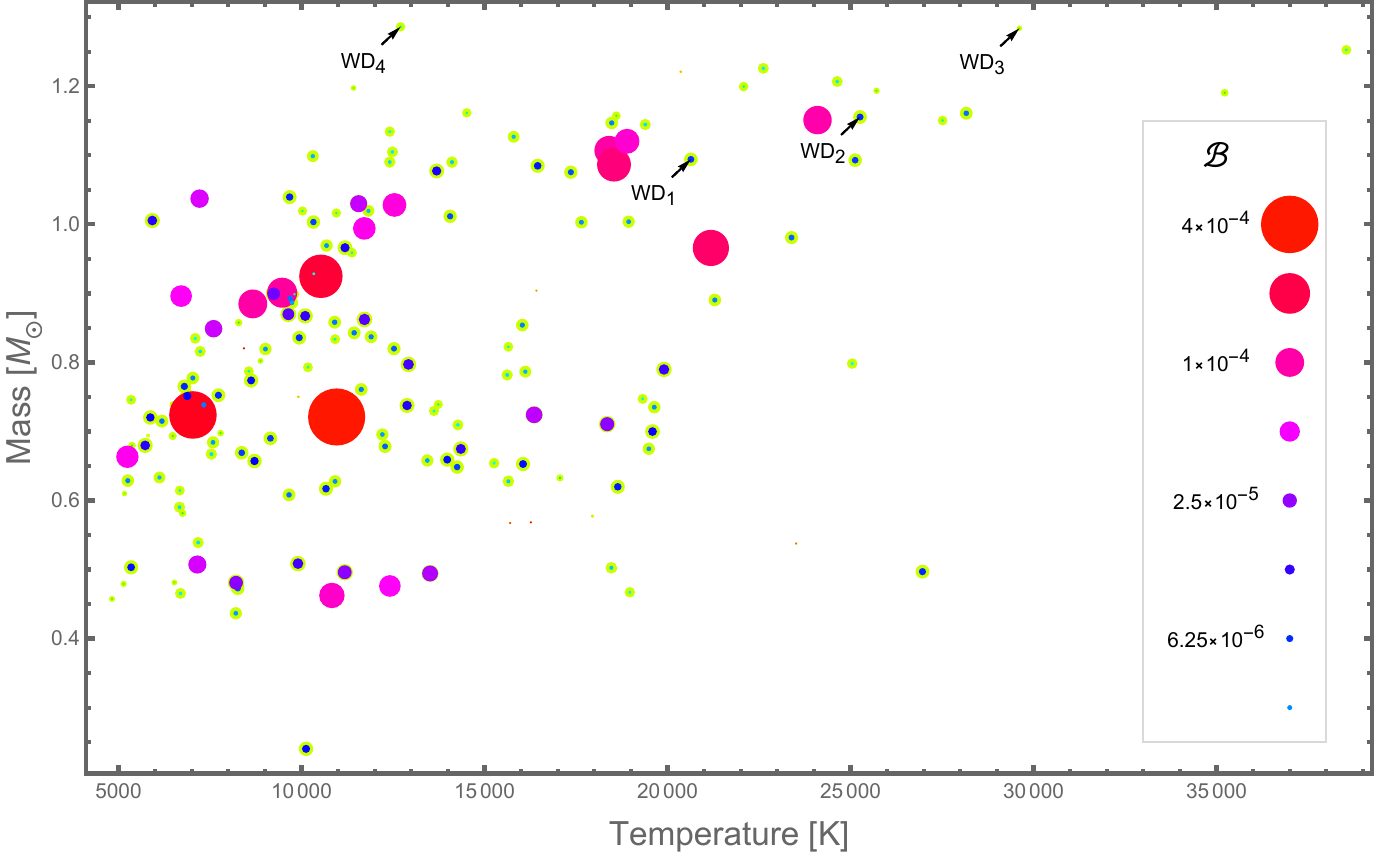}
             {\caption{Mass, temperature and magnetic parameter ${\cal B}$ of magnetic white dwarfs from Gentile's collection, with arrows pointing to the four with known rotation rate from \citep{2024ApJ...974...12J}. Magnetic parameter is coded by size and color as shown in inset; stars with ${\cal B}<3\times 10^{-4}$ are represented by pale greenish dots.} \label{BelePritl}}
    \end{figure*}

White dwarfs are known to be magnetic and do rotate. However, less data give their measured mass together with rotation period and magnetic field. Two such examples are white dwarfs   J221141.80 \citep{2021ApJ...923L...6K} and J1745 \citep{2023ApJ...954..138V} with known rotation rates (P=614$s$,  M=0.92$M_\odot$, $B>7\times 10^6$gauss) and (P=70.36, M=1.268$M_\odot$, $B=15\times 10^6$gauss)\footnote{\label{FTNwdMR}Since  white dwarf radii are not given in the quoted sources, we use the simple mass-radius relation $R_{wd}\sim \frac{\hbar^2}{2m_e G\,M^{1/3}(Z\,m_p)^{5/3}}$ with $Z=1/2$ (number of electrons per nucleon)}  are shown in Fig.\ref{SLRotMagEn} as small blue spheres  numbered 1 and 2. Magnetic field can also be detected on some white dwarfs through Zeeman splitting of hydrogen and helium lines (\cite{2023ApJ...944...56A},\cite{2017ASPC..509....3D},\cite{2023MNRAS.520.6135H},\cite{2023MNRAS.520.6111H}), however data on their rotation are generally not available. Nicola Pietro Gentile Fusillo from University of Trieste has been kind enough to provide a private collection of available data containing information on mass, temperature, parallax, spectral data and magnetic field of 139 white dwarfs. In Fig.\ref{BelePritl} we display these data as the distribution of  magnetic white dwarfs with respect to mass and temperature\footnote{The magnetic parameter $\cal B$ was calculated as above, cf. footnote \ref{FTNwdMR} }.   It was quite surprising to find out that white dwarfs are distributed on the same interval of magnetic moment $\cal B$ as all other celestial objects shown in Fig.\ref{SLRotMagEn}. A recent paper \citep{2024ApJ...974...12J} quotes rotational periods of a number of magnetic white dwarfs, four of which ($\mathrm {WD}_1$: J0043-1000,  $\mathrm {WD}_2$: J1214-1724, $\mathrm {WD}_3$: J1659-4401 and $\mathrm {WD}_4$: J2257-0755) are also members of Nicola's collection. They are pointed at by arrows in Fig.\ref{BelePritl} and are also introduced in Fig.\ref{SLRotMagEn} as  blue balls. A recent article \citep{2024MNRAS.528.6056H} quotes rotation periods masses and magnetic field  of ten more white dwarfs (WD1: 0011-134, WD 0009+501, WD 2359-434, WD 0011-721, WD J075328.47-511436.98, WD J171652.09-590636.29,  WD 0912+536, WD 0041-102, WD 2138-332, LSPM J0107+2650). They are represented by  blue balls and are numbered from 3 to 12 in Fig.\ref{SLRotMagEn}.

A recently studied Wolf Rayet star HD 45166 with very high magnetic field (\cite{2023Sci...381..761S}) is also shown in  Fig.\ref{SLRotMagEn} as a blue ball. The high  $\cal B$ value of this star with respect to white dwarfs is due to the fact that HD is a ``normal size''  star  ($R=2.6\,R_\odot,\,\, M=3.4\,M_\odot$), while  white dwarfs are more than 100 times smaller in radius, therefore, their $\cal B$ value (which measures magnetic flux per mass of the star) may be less even if their magnetic field is higher.  
\section{Discussion}
The phase diagram in Fig.\ref{SLRotMagEn} may be considered as  a demonstration of interplay between magnetism and rotation in breaking the symmetry of gravity. We find 
that very different celestial objects, planets, normal stars, white dwarfs and pulsars occupy more or less the same, relatively limited region in this diagram, where angular velocity ($\tilde \omega$) is scaled by characteristic rotational breakup velocity ($\Omega$) and magnetic field parameter $\cal B$ representing magnetic flux  per mass of the star. On a linear scale this region would appear as an extremely narrow band, only
about two orders of magnitude wide but extending over nearly eight orders of
magnitude in length, which makes the observed clustering even more striking.

Remembering that  rotational and  magnetic dipole and total energy of a star can be expressed as  $W_{rot}\sim \frac{1}{5}M\,a^2\,\Omega^2\,{\tilde \omega}^2$, $W_{mag}\sim \frac{1}{2}M\,a^2\,\Omega^2\,{\cal B}^2$,  and  $W_{total}\sim-\frac{1}{3}M\,a^2\,\Omega^2$, it can be understood that the phase diagram  represents a correlation between magnetic, rotational  and  total energy of stars.  
In this respect we note that for every white dwarf we can find a nearby pulsar in the diagram, for example 2 and Crab, 3 and Vela etc. It is also interesting that white dwarfs (9) can be found near the pulsar-magnetar division and white dwarfs 7, 6 and 11 are very close to pulsar's death line.  


 Fig.\ref{SLRotMagEn} also suggests that slower (older?) pulsars have stronger magnetic field then fast ones. This appearance  is usually explained by the fact that  pulsars reaching the death line (marked in magenta) become unobservable, while fast pulsars with strongest magnetic field (marked in brown) slow down so fast that few are observed both with high magnetic field and fast rotation. However, if a similar correlation between high magnetic field and slow rotation persists in the case of white dwarf population, the above argument can not be applied to justify the shortage of slow low magnetic field white dwarfs. Firstly, the magnetic field of white dwarfs is not deduced from their slow-down rate and secondly, the characteristic decay time of white dwarf slow down due to magnetic dipole radiation is incomparably long with respect to pulsars\footnote{in terms of dimensionless  ($\tilde \omega$) and  ($\cal B$), the radiation braking decay time can be expressed as: $\tau_d=\frac{\omega}{\dot \omega}=\frac{4}{5}\frac{a}{c}\left(\frac{a\,c^2}{G M}\right)^2\frac{1}{{\cal B}^2{\tilde \omega}^2}$, so it is roughly $\sim (100)^3$ times longer for  white dwarfs.}. 
 
 
 
We also note that our simple model requires a rederivation ot the slow-down law for pulsars, since the energy lost by dipole radiation is released by all available energy reservoirs. The fundamental law to be taken into account is the law of conservation of angular momentum stating $\frac{dL}{dt}=-\frac{P_{rad}}{\omega}$, where $L$ is the angular momentum of the star and $P_{rad}=\frac{1}{12\pi}\sqrt{\frac{\mu_0}{\varepsilon_0}}\left(\frac{\omega}{c}\right)^4{\cal M}^2$ is the dipole radiation power. Expressing magnetic moment ($\cal M$) and angular velocity $\omega$ with dimensionless quantites $\cal B$ and $\tilde \omega$, the dipole radiation power becomes: $P_{rad}=\frac{1}{3c^3}a^2 G\,M^2 \Omega^4 {\cal B}^2{\tilde \omega}^4$. Expressing the angular momentum as $L=\frac{2W_{rot}}{\omega}$, one obtains $\frac{dL}{dt}=\frac{2}{5}M\,a^2\Omega^2(1+e^2)\left[(1+e^2)\frac{d{\tilde\omega}}{dt}+4{\tilde \omega}\,e\frac{de}{dt}\right]$. The term $e\frac{de}{dt}$ can be expressed as $e\,de\approx \frac{15}{4}\,{\tilde \omega}\,d{\tilde\omega}+\frac{81}{4}{\cal B}\,d{\cal B}$ by taking the total derivative of equation  (\ref{BboundCRot}) if terms of higher order in $\cal B$ and $\tilde\omega$ are neglected. Obviously $\frac{d\tilde\omega}{dt}\gg {\tilde\omega}\,e\,\frac{de}{dt}$, so to first order
the angular momentum conservation law leads to $\frac{d{\tilde \omega}}{dt}=-\frac{5}{6(1+e^2)^2}\frac{G\,M\,\Omega^2}{c^3}{\cal B}^2{\tilde \omega}^3$, which is the classical result, except for the fact that $e$ is a function of time. Taking this into account in calculating the second derivative of rotational frequency, we obtain:$\frac{d^2{\tilde \omega}}{dt^2}=2\frac{d{\tilde\omega}}{dt}\left(\frac{d{\cal B}}{{\cal B}\,dt}+\frac{3}{2}\frac{d\tilde\omega}{{\tilde \omega}dt}\right)$. The braking index ($n_b=\frac{{\tilde\omega}\,\ddot{\tilde\omega}}{{\tilde\omega}^2}$) so becomes\footnote{Having above neglected  terms of higher order, we understand $1+e^2\sim 1$}: $n_b\approx 3-\left(\frac{12}{5}\frac{c^3}{G\,M\,\Omega^2}\frac{1}{{\tilde\omega}^2{\cal B}^3}\frac{d{\cal B}}{dt}+15{\tilde\omega}^2\right)$.
 
The discussion so far is based on the simplest $n=0$ polytropic model, assuming 
that the magnetic field deduced from stellar spectra and polarization properties applies to the observed dipole component. Since this model deals with incompressible fluid, the magnetic surface can only be at the surface of the star, which results in  very strong coupling between  gravity and magnetism. As a result of such strong coupling the magnetic surface instability sets in only for very high order spherical harmonics expansion.  

 This assumption may seem acceptable regarding magnetic fields of old neutron stars, white dwarfs and planets, but less so regarding the solar magnetic field, which appears  in the above phase diagram much too  weak to have any effect on oblateness. Yet the solar oblateness  \citep{2018csc..confE...5B}  is too small even  with respect to the softest polytropic model, while \citep{2008Sci...322..560F} reported a slight variation that could be correlated with solar  activity.  Furthermore, the solar magnetic field is characterized by spikes of intense magnetic field in sunspots \citep{2018ApJ...852L..16O}, which may be reminiscent of magnetic surface instability  forming  ''magnetic hair" on magnetic surface, deep below the photosphere. Convection, differential rotation and meridional flow may tangle magnetic hair appearing at sunspots and drive solar magnetic activity. 
In this respect the magnetic Wolf Rayet star HD 45166 with magnetic field of 43kG is interesting. WR stars are known as stars stripped of their envelope. Noting that the magnetic field 43kG is less than 10 times stronger then the magnetic field in sunspots \citep{2018ApJ...852L..16O}, one may ask if this is an indication, that HD 45166 was stripped down to its magnetic surface.

The wide variety of pulsar's pulse shapes at high energies \citep{2009ApJ...696.1084A} also  suggests that the dipole field alone can not be responsible for sharp edges of gamma ray pulses. In particular, the detailed study of Vela's gamma-- ray pulse at  energies from 50$\,\mathrm{MeV}$ to 30$\,\mathrm{GeV}($ \cite{2011ASSP...21...37R},\cite{2010ApJ...713..154A}) reveals the increasing complexity of pulse shape with increasing energy, which is consistent with the idea that the high-- energy pulse is generated closer to the pulsar surface where the higher multipole order of the magnetic field dominates over the dipole. The curious pulse structure of intermittend $10.4\,\, s$ pulsar PSR $\mathrm  J1710-3452$ (\cite{2023MNRAS.526L.143S}), with a magnetic field possibly as much as $10^4$ times as strong as the Crab, adds to the mistery of the multipole structure of highly magnetized~stars. 

\section{Conclusion}

This work was stimulated by a simple question: what is the difference, appart from scale, between magnetism of a permanent magnet and the magnetism of a star?  

According to everyday experience, matter needs to be magnetized by driven currents to become magnetic. Yet, once magnetized, a magnet  as well as a star persist in beeing magnetic indefinitely. This  suggests that the state of beeing magnetic is a characteristic of equilibrium of a material body with its electro-magnetic environment. In a similar way, spin may be charaterizing the way in which a body is imbedded into its surroundings. Therefore, we attempt to include rotation and  electromagnetic interaction in the basic discussion of stellar structure. 

The starting point of this endeavour is the polytropic model which predicts the mass-radius relation for a celestial body based on the law of gravitation and a rather simple equation of state for stellar material. We formulate the solution to this problem through a variational principle and extend it to include also  spin  and magnetic moment as two additional observable parameters.  

The emergence of magnetic moment is demonstrated by deriving the equation of state for degenerate matter consisting of cold heavy ions and the Fermi gas of light electrons.  We show that such a quantum system  has a magnetized ground state.  A similar conclusion, yet by different means, was obtained by \cite{2022PNAS..11919831Z} in their study of spontaneous magnetization of collisionless plasma\footnote{The  requirement that electrons are moving almost umperturbed in the electromagnetic environment is the common feature needed in both derivations to obtain the magnetic ground state.}.

An exact solution of the extended gravito-magnetic polytropic problem was found for the simplest polytropic model with $n=0$. Models with a soft equation of state are much more demanding, since the magnetic contribution to action is formulated as a surface integral over a generally unstable magnetic surface, which lies below the physical surface of the star.  The instability of magnetic surface is briefly discussed.
\vspace{0.5 cm}

 In  Fig.\ref{SLRotMagEn} we distribute stars with available data on their rotation and magnetic field  in a phase diagram based on the exact $n=0$ polytropic solution. The axes of this diagram are magnetic parameter ($\cal B$), the spin parameter ($\tilde \omega$) and energy excess $W$.   $\cal B$ represents the ratio of magnetic flux through the equatorial section and mass of the star,   ($\tilde \omega$) is the ratio of the measured rotational velocity of the star and the breakup velocity $\Omega=\sqrt{2\pi G \rho}$ and $W$ is the free energy of the rotating magnetic star with respect to the free energy of non-rotating and no-magnetic star. It is quite remarkable that so many different celestial objects occupy the same region in this diagram. We believe that this coincidence supports the idea that electromagnetic field excercises its long--range character through magnetism.





\begin{thebibliography}{0}%
\makeatletter
\providecommand \@ifxundefined [1]{%
 \@ifx{#1\undefined}
}%
\providecommand \@ifnum [1]{%
 \ifnum #1\expandafter \@firstoftwo
 \else \expandafter \@secondoftwo
 \fi
}%
\providecommand \@ifx [1]{%
 \ifx #1\expandafter \@firstoftwo
 \else \expandafter \@secondoftwo
 \fi
}%
\providecommand \natexlab [1]{#1}%
\providecommand \enquote  [1]{``#1''}%
\providecommand \bibnamefont  [1]{#1}%
\providecommand \bibfnamefont [1]{#1}%
\providecommand \citenamefont [1]{#1}%
\providecommand \href@noop [0]{\@secondoftwo}%
\providecommand \href [0]{\begingroup \@sanitize@url \@href}%
\providecommand \@href[1]{\@@startlink{#1}\@@href}%
\providecommand \@@href[1]{\endgroup#1\@@endlink}%
\providecommand \@sanitize@url [0]{\catcode `\\12\catcode `\$12\catcode
  `\&12\catcode `\#12\catcode `\^12\catcode `\_12\catcode `\%12\relax}%
\providecommand \@@startlink[1]{}%
\providecommand \@@endlink[0]{}%
\providecommand \url  [0]{\begingroup\@sanitize@url \@url }%
\providecommand \@url [1]{\endgroup\@href {#1}{\urlprefix }}%
\providecommand \urlprefix  [0]{URL }%
\providecommand \Eprint [0]{\href }%
\providecommand \doibase [0]{https://doi.org/}%
\providecommand \selectlanguage [0]{\@gobble}%
\providecommand \bibinfo  [0]{\@secondoftwo}%
\providecommand \bibfield  [0]{\@secondoftwo}%
\providecommand \translation [1]{[#1]}%
\providecommand \BibitemOpen [0]{}%
\providecommand \bibitemStop [0]{}%
\providecommand \bibitemNoStop [0]{.\EOS\space}%
\providecommand \EOS [0]{\spacefactor3000\relax}%
\providecommand \BibitemShut  [1]{\csname bibitem#1\endcsname}%
\let\auto@bib@innerbib\@empty
\end{thebibliography}%


\begin{thebibliography}{99}
\bibitem[Abdo et al.(2009)]{2009ApJ...696.1084A} Abdo, A.~A., Ackermann, M., Atwood, W.~B., et al.\ 2009, \apj, 696, 1084. doi:10.1088/0004-637X/696/2/1084
\bibitem[Abdo et al.(2010)]{2010ApJ...713..154A} Abdo, A.~A., Ackermann, M., Ajello, M., et al.\ 2010, \apj, 713, 154. doi:10.1088/0004-637X/713/1/154
\bibitem[Amorim et al.(2023)]{2023ApJ...944...56A} Amorim, L.~L., Kepler, S.~O., K{\"u}lebi, B., et al.\ 2023, \apj, 944, 56. doi:10.3847/1538-4357/acaf6e
\bibitem[Burgio \& Vida{\~n}a(2020)]{2020Univ....6..119B} Burgio, G.~F. \& Vida{\~n}a, I.\ 2020, Universe, 6, 119. doi:10.3390/universe6080119
\bibitem[Bush et al.(2018)]{2018csc..confE...5B} Bush, R.~I., Emilio, M., Scholl, I., et al.\ 2018, Catalyzing Solar Connections, 5 
\bibitem[Cadez \& Javornik(1981)]{1981Ap&SS..77..299C} \v Cade\v z, A. \& Javornik, M.\ 1981, \apss, 77, 299. doi:10.1007/BF00649461
\bibitem[{\v{C}}ade{\v{z}} et al.(2016)]{2016A&A...587A..99C} {\v{C}}ade{\v{z}}, A., Zampieri, L., Barbieri, C., et al.\ 2016, \aap, 587, A99. doi:10.1051/0004-6361/201526490
\bibitem[Caleb et al.(2022)]{2022NatAs...6..828C} Caleb, M., Heywood, I., Rajwade, K., et al.\ 2022, Nature Astronomy, 6, 828. doi:10.1038/s41550-022-01688-x
\bibitem[Caleb et al.(2024)]{2024NatAs...8.1159C} Caleb, M., Lenc, E., Kaplan, D.~L., et al.\ 2024, Nature Astronomy, 8, 1159. doi:10.1038/s41550-024-02277-w 
\bibitem[Dufour et al.(2017)]{2017ASPC..509....3D} Dufour, P., Blouin, S., Coutu, S., et al.\ 2017, 20th European White Dwarf Workshop, 509, 3. doi:10.48550/arXiv.1610.00986
\bibitem[Event Horizon Telescope Collaboration et al.(2024)]{2024ApJ...964L..26E} Event Horizon Telescope Collaboration, Akiyama, K., Alberdi, A., et al.\ 2024, \apjl, 964, L26. doi:10.3847/2041-8213/ad2df1
\bibitem[Fivian et al.(2008)]{2008Sci...322..560F} Fivian, M.~D., Hudson, H.~S., Lin, R.~P., et al.\ 2008, Science, 322, 560. doi:10.1126/science.1160863
\bibitem[Hardy et al.(2023)]{2023MNRAS.520.6135H} Hardy, F., Dufour, P., \& Jordan, S.\ 2023, \mnras, 520, 6135. doi:10.1093/mnras/stad197
\bibitem[Hardy et al.(2023)]{2023MNRAS.520.6111H} Hardy, F., Dufour, P., \& Jordan, S.\ 2023, \mnras, 520, 6111. doi:10.1093/mnras/stad196
\bibitem[Harwit(2006)]{2006asco.book.....H} Harwit, M.\ 2006, . doi:10.1007/978-0-387-33228-4
\bibitem[Hernandez et al.(2024)]{2024MNRAS.528.6056H} Hernandez, M.~S., Schreiber, M.~R., Landstreet, J.~D., et al.\ 2024, \mnras, 528, 6056. doi:10.1093/mnras/stae307
\bibitem[Jewett et al.(2024)]{2024ApJ...974...12J} Jewett, G., Kilic, M., Bergeron, P., et al.\ 2024, \apj, 974, 12. doi:10.3847/1538-4357/ad6905
\bibitem[Kittel(1976)]{1976itss.book.....K}
Kittel, C.\ 1976, \textit{Introduction to Solid State Physics}, 5th ed., Wiley.
\bibitem[K{\"u}lebi et al.(2009)]{2009A&A...506.1341K} K{\"u}lebi, B., Jordan, S., Euchner, F., et al.\ 2009, \aap, 506, 1341. doi:10.1051/0004-6361/200912570
\bibitem[Okamoto \& Sakurai(2018)]{2018ApJ...852L..16O} Okamoto, T.~J. \& Sakurai, T.\ 2018, \apjl, 852, L16. doi:10.3847/2041-8213/aaa3d8
\bibitem{MTW} C.~W.~Misner, K.~S.~Thorne, and J.~A.~Wheeler, \textit{Gravitation}, W.~H.~Freeman, 1973.
\bibitem[Monaghan \& Roxburgh(1965)]{1965MNRAS.131...13M} Monaghan, J.~J. \& Roxburgh, I.~W.\ 1965, \mnras, 131, 13. doi:10.1093/mnras/131.1.13
\bibitem[Monaghan(1965)]{1965MNRAS.131..105M} Monaghan, J.~J.\ 1965, \mnras, 131, 105. doi:10.1093/mnras/131.1.105 
\bibitem[Irbah et al.(2019)]{2019ApJ...875L..26I} Irbah, A., Mecheri, R., Dam{\'e}, L., et al.\ 2019, \apjl, 875, L26. doi:10.3847/2041-8213/ab16e2
\bibitem[Kilic et al.(2021)]{2021ApJ...923L...6K} Kilic, M., Kosakowski, A., Moss, A.~G., et al.\ 2021, \apjl, 923, L6. doi:10.3847/2041-8213/ac3b60
\bibitem[Rosensweig(1987)]{1987AnRFM..19..437R} Rosensweig, R.~E.\ 1987, Annual Review of Fluid Mechanics, 19, 437. doi:10.1146/annurev.fl.19.010187.002253
\bibitem[Ray \& Parkinson(2011)]{2011ASSP...21...37R} Ray, P.~S. \& Parkinson, P.~M.~S.\ 2011, High-Energy Emission from Pulsars and their Systems, 21, 37. doi:10.1007/978-3-642-17251-9\_3
\bibitem[Shenar et al.(2023)]{2023Sci...381..761S} Shenar, T., Wade, G.~A., Marchant, P., et al.\ 2023, Science, 381, 761. doi:10.1126/science.ade3293
\bibitem[Sharma et al.(2023)]{2023arXiv231004079S} Sharma, R., Jain, C., Paul, B., et al.\ 2023, arXiv:2310.04079. doi:10.48550/arXiv.2310.04079
\bibitem[Sob'yanin(2023)]{2023PhRvD.107h1301S} Sob'yanin, D.~N.\ 2023, \prd, 107, L081301. doi:10.1103/PhysRevD.107.L081301
\bibitem[Surnis et al.(2023)]{2023MNRAS.526L.143S} Surnis, M.~P., Rajwade, K.~M., Stappers, B.~W., et al.\ 2023, \mnras, 526, L143. doi:10.1093/mnrasl/slad082
\bibitem[Vermette et al.(2023)]{2023ApJ...954..138V} Vermette, B., Salcedo, C., Mori, K., et al.\ 2023, \apj, 954, 138. doi:10.3847/1538-4357/ace90c
\bibitem[Valyavin et al.(2005)]{2005A&A...439.1099V} Valyavin, G., Bagnulo, S., Monin, D., et al.\ 2005, \aap, 439, 1099. doi:10.1051/0004-6361:20052642\%white dwarf 0009+501
\bibitem[Zhou et al.(2022)]{2022PNAS..11919831Z} Zhou, M., Zhdankin, V., Kunz, M.~W., et al.\ 2022, Proceedings of the National Academy of Science, Spontaneous magnetization of collisionless plasma, 119, 19, e2119831119. doi:10.1073/pnas.2119831119
\bibitem[Bruhat et al.(1996)]{Cecile1996}Yvonne Choquet-Bruhat, Cecile DeWitt-Morette, Margaret Dillard-Bleick: Analysis, Manifolds and Physics 
\end{thebibliography}

\end{document}